\documentclass[aps,prc,tightenlines,twoside,twocolumn,fleqn,superscriptaddress]{revtex4-1}

\usepackage{amsmath,amsthm,amssymb}
\usepackage{graphicx}
\usepackage{soul}
\usepackage{color}
\usepackage{xcolor}

\colorlet{darkgreen}{green!50!black}
\colorlet{brightyellow}{yellow!75!red}
\colorlet{orange}{red!50!yellow}
\colorlet{darkblue}{blue!60!black}
\colorlet{darkred}{red!80!black}

\newcommand{\dif}{\mathrm{d}}
\begin{document}

\title{Initial Angular Momentum and Flow in High Energy Nuclear Collisions}



\author{Rainer J.\ Fries}
\email{rjfries@comp.tamu.edu}
\affiliation{Cyclotron Institute and Department of Physics and Astronomy, Texas A\&M University, College Station TX 77843, USA}
\affiliation{Department of Physics, McGill University, 3600 University Street, Montreal, QC, H3A 2T8, Canada}

\author{Guangyao Chen} 
\email{gchen@iastate.edu}
%\thanks{}
\affiliation{Department of Physics and Astronomy, Iowa State University, Ames IA 50011, USA}
\affiliation{Cyclotron Institute and Department of Physics and Astronomy,
  Texas A\&M University, College Station TX 77843, USA}

\author{Sidharth Somanathan}
\affiliation{Cyclotron Institute and Department of Physics and Astronomy, Texas A\&M University, College Station TX 77843, USA}

\date{\today}

\begin{abstract}
We study the transfer of angular momentum in high energy nuclear collisions from 
the colliding nuclei to the region around midrapidity, using the classical approximation 
of the Color Glass Condensate (CGC) picture. We find that the angular momentum 
shortly after the collision (up to times $\sim 1/Q_s$, where $Q_s$ is the saturation scale)
is carried by the ``$\beta$-type" flow of the initial classical gluon field, introduced by some of us earlier. 
$\beta^i \sim  \mu_1 \nabla^i\mu_2 - \mu_2\nabla^i \mu_1$ ($i=1,2$) 
describes the rapidity-odd transverse energy flow and emerges from Gauss' Law for 
gluon fields. Here $\mu_1$ and $\mu_2$ are the averaged color charge fluctuation densities
in the two nuclei, respectively. 
Interestingly, strong coupling calculations using AdS/CFT techniques
also find an energy flow term featuring this particular combination of nuclear densities.
In classical CGC the order of magnitude of the initial angular 
momentum per rapidity in the reaction plane, at a time $1/Q_s$,
is $|dL_2/d\eta| \approx  R_A Q_s^{-3} \bar\varepsilon_0/2 $ at midrapidity,
where $R_A$ is the nuclear radius, and $\bar\varepsilon_0$ 
is the average initial energy density.
This result emerges as a cancellation between a vortex of energy flow in the reaction plane aligned with
the total angular momentum, and energy shear flow opposed to it.
We discuss in detail the process of matching classical Yang-Mills 
results to fluid dynamics. We will argue that dissipative corrections should not be discarded
to ensure that macroscopic conservation laws, e.g.\ for angular momentum, hold.
Viscous fluid dynamics tends to dissipate the shear flow contribution that carries angular momentum
in boost-invariant fluid systems. This leads to small residual angular momentum around midrapidity at late times for collisions 
at high energies.
\end{abstract}

\pacs{24.85.+p,25.75.Ag,25.75.Ld}
\keywords{Heavy Ion Collisions, Color Glass Condensate, Collective Flow}

\maketitle

\section{Introduction}

Angular momentum carried by the hot nuclear matter produced in heavy ion collisions has 
received a great deal of attention in recent years. This has been amplified by the announcement 
of the STAR experiment at the Relativistic Heavy Ion Collider (RHIC) that $\Lambda$ and 
$\bar\Lambda$ baryons show a noticeable amount of polarization \cite{STARLambda}, consistent
with the direction of total angular momentum of the system.
This polarization is small or consistent with zero at top RHIC energies but rises significantly 
towards lower beam energies. This behavior as a function of energy can be understood, qualitatively, 
as a suppression of shear flow around midrapidity in systems with increasingly good boost-symmetry, as we 
will discuss below.

In the literature the production of polarized particles, and fluid dynamics with vorticity,
%, and possible initial conditions that can lead to finite angular momentum 
have been discussed extensively
\cite{Liang:2004ph,Becattini:2013vja,Csernai:2013bqa,Becattini:2013vja,Jiang:2016woz,Pang:2016igs,Deng:2016gyh,Xie:2017upb}.
Here we try to address the simple question how angular momentum of the
colliding nuclei is transferred toward midrapidity in the initial phase of the collision, therefore possibly seeding
polarization effects at later stages. We focus on high energy collisions where longitudinal boost-invariance holds 
close to midrapidity, and where Color Glass Condensate (CGC) \cite{McLerran:1993ka,McLerran:1993ni, JMKMW:96,
  KoMLeWei:95,Kovner:1995ja,Kovchegov:1996ty,Kovchegov:1998bi,
  Iancu:2003xm,Gelis:2010nm} is thought to be the correct effective theory for the initial phase 
of the nuclear interaction. Initial conditions with angular momentum based on transport models or similar considerations,
appropriate for smaller collision energies, have been discussed in the literature, e.g.\  in 
\cite{Becattini:2007sr,Csernai:2013bqa,Becattini:2016gvu,Karpenko:2016jyx,Li:2017dan}.
The advantage of the Color Glass Condensate approach lies in the fact that it should be the correct effective
theory at asymptotically large collision energies. Moreover, in the classical Yang-Mills approximation to CGC one can 
derive analytic estimates for the angular momentum in the system at early times, based on previous work 
\cite{Fries:2006pv,Chen:2013ksa,Chen:2015wia}. We show that at the highest collision energies angular momentum 
is rapidly built up around midrapidity by Gauss' Law in the classical gluon field phase.

In the CGC approach strong quasi-classical gluon fields dominate the initial
interaction of soft and semi-soft modes, characterized by the saturation scale
$Q_s$. Such modes after the collision are generated from similar modes 
present in the wave function of the nuclei before the collision. The purely classical description of CGC, 
featuring a Gaussian sampling of color charge fluctuations, is known as the McLerran-Venugopalan (MV) 
model and will be used here \cite{McLerran:1993ka,McLerran:1993ni}. 
In this work we will focus on event-averaged quantities and work in a strictly boost-invariant setup so 
that the classical Yang-Mills equations can be solved with the recursion relations from Refs.\ 
\cite{Fries:2006pv,Chen:2013ksa,Chen:2015wia}. Enforcing boost-invariance limits the applicability to the largest
collision energies. We will also restrict ourselves to a few lowest orders of the recursive solution
which limits our reach in time. We refer the reader to Ref.\ \cite{Li:2016eqr} for a 
recent attempt at resummation of the recursive solution. Our simple approach here will allow us to derive a 
pocket formula that relates the angular momentum at midrapidity to the initial energy density at a characteristic time 
scale $\tau = 1/Q_s$. In the future our analytic results could be tested against, and refined by, numerical simulations of the classical Yang-Mills system.

The validity of the purely classical description of the gluon field ceases around a time $\sim 1/Q_s$, where decoherence 
and the growth of fluctuations \cite{Dusling:2011rz} set the system on a path towards thermalization 
\cite{Berges:2013eia,Berges:2013fga}. Eventually quark gluon plasma (QGP) is created close to local kinetic equilibrium. 
Hybrid calculations that use the MV model and match the results directly to viscous fluid dynamics 
\cite{Israel:1979wp,Muronga:2004sf,Heinz:2005bw,Romatschke:2007mq,Dusling:2007gi,Schenke:2010nt,Karpenko:2013wva}, without dynamical 
simulations of the thermalization process, have been phenomenologically very successful, pioneered by the 
IP-Glasma + MUSIC approach \cite{Schenke:2012wb,Schenke:2012fw,Gale:2012rq}. 
Following this example we match the energy momentum tensor of the classical Yang-Mills field
at a time $\tau_0$ around $1/Q_s$ directly to fluid dynamics (see \cite{vanderSchee:2013pia}
for a similar procedure in a strong coupling scenario). We discuss the merits and limitations of this procedure.
We argue that it is important to keep dissipative stress computed from the Yang-Mills fields at the matching 
time in order to ensure that the conservation laws for energy, momentum and angular momentum are obeyed.
However, even in that case important physics is missing around the matching time and time derivatives can change sign rapidly.

We further trace angular momentum through the subsequent 
viscous fluid evolution. Dissipative corrections directly emerging from a system of classical 
Yang-Mills fields are expected to be large, and conventional viscous fluid dynamics is not necessarily 
well equipped to handle large corrections in a reliable way. This could in the future be improved by 
matching to anisotropic fluid dynamics \cite{Martinez:2010sc,Bazow:2013ifa}. In either case
some of the equations of motion sharply change at the matching point, and for some quantities the time evolution 
before and after the matching time $\tau_0$ can be very different. 
The flow component carrying angular momentum is an important example. As discussed in detail below, 
shear flow is built up in the Yang-Mills phase due to Gauss' Law. However, in the viscous fluid phase the 
Navier-Stokes mechanism provides a damping effect which decreases shear flow. 
A more complete microscopic description of the thermalization regime, starting from the classical CGC phase
\cite{Gelis:2008rw,Berges:2013eia,Berges:2013fga}, should smoothen the transition.

Thus, while in the classical field phase angular momentum is actively built up around midrapidity, 
the opposite is true in the viscous fluid phase. The flow field of a boost-invariant fluid can carry 
angular momentum only through longitudinal shear flow, and thus viscosity effects diminish the angular 
momentum at midrapidity. 
Longitudinal shear flow is also a general aspect of other initial state models that do not have boost-invariance
\cite{Becattini:2007sr,Csernai:2013bqa}. Without boost-invariance, global rotation of the system becomes 
an increasingly effective option to carry angular momentum 
\cite{Becattini:2007sr,Jiang:2016woz,Csernai:2015jsa}.

Briefly returning to the limiting assumptions in this paper, we note that boost-invariance allows us
to make statements only about nuclear collision systems at the largest energies, and also then only for the part
of the system away from beam rapidities. Collisions at top RHIC and LHC energies are examples of systems where
this approximation is probably meaningful. We also integrate out transverse fluctuations and disallow (by boost symmetry)
longitudinal fluctuations. The transverse integral limits us to make statements about event-averaged quantities and we
leave event-by-event fluctuations to a future publication. The absence of longitudinal fluctuations limits us to times 
$\lesssim 1/Q_s$ as they can grow very large at later times. Despite these limitations we establish benchmarks
in this paper that future calculations can be checked against.

The paper is organized as follows. In the next section we review
some basic results about the classical gluon field in high energy nuclear collisions.
We analyze the resulting angular momentum tensor and obtain expressions
for the angular momentum per rapidity in the reaction plane. In section 
\ref{sec:match} we discuss a suitable matching procedure between classical fields
and fluid dynamics, first in the ideal case, then in the viscous case, using 
conservation laws as the guiding principle. In section \ref{sec:fluid}  we analyze the 
initial fluid system obtained by matching to the classical gluon fields. We focus 
on the components of the velocity field and the viscous shear stress tensor that 
contribute to the angular momentum in the reaction plane.
We follow up with a time evolution of the system using the VIRAL viscous fluid code. 
The paper concludes with a summary and discussion.

\section{Analytic Solutions for Color Glass Condensate}
\label{sec:2}

\subsection{A Review of Previous Results}

In the MV realization of color glass a nucleus is represented by a current 
of color charge on the light cone which creates a classical gluon field. 
Hence a collision of nuclei is set up by two opposing components of
light cone currents,
\begin{align}
  \label{eq:current}
  J^+ (x) &= \delta(x^-) \rho_1(\vec x_\perp)\,  ,
  \\
  J^-_2(x) &= \delta(x^+) \rho_2(\vec x_\perp)\, , \\
  J^i_{1,2} (x) &= 0 \, ,
\end{align}
with $i=1,2$, where $\rho_1(\vec x_\perp)$ and $\rho_2(\vec x_\perp)$ are the 
transverse densities of color charge in nucleus $1$ and $2$, respectively. Light cone coordinates 
are defined as $x^\pm = (x^0\pm x^3)/\sqrt{2}$. This current
satisfies the continuity equation $D_\mu J^\mu=0$ if we choose an axial gauge with
\begin{equation}
  x^+ A^- + x^- A^+ = 0,
\end{equation}
and the gluon field $F^{\mu\nu} = i[D^\mu,D^\nu]/g$, with covariant
derivative $D^\mu =\partial^\mu -ig
A^\mu$ is generated by the color current through the Yang-Mills equation
\begin{equation}
  \left[ D_\mu, F^{\mu\nu}\right] = J^\nu  \, .
  \label{eq:ym}
\end{equation}
Note that the current is manifestly boost-invariant. However, because the two
charges $\rho_1$ and $\rho_2$ will generally not be the same, there is no symmetry under
interchange of the $+$ and $-$ direction.

The gluon field that forms after the collision in the forward light cone 
can be solved numerically 
\cite{Krasnitz:2000gz,Lappi:2003bi,Krasnitz:2003jw,Schenke:2012wb}, or
analytically using a power series in proper time $\tau=\sqrt{t^2-z^2}$
\cite{Fries:2006pv,Fries:2005yc,Chen:2013ksa,Chen:2015wia}. We will now
review some of the results obtained through the analytic approach.
All quantities are written as a power series in the forward light cone
($\tau \ge 0$), for example the energy momentum tensor
\begin{equation}
  T^{\mu\nu}_{\mathrm{YM}} (\tau,\vec x_\perp) = \sum_{n=0}^\infty \tau^n
  T^{\mu\nu}_{(n)}(\vec x_\perp) \, .
\end{equation}
The Yang-Mills equations can then be solved recursively with boundary
conditions at $\tau=0$. These boundary conditions lead to
longitudinal chromo-electric and -magnetic fields
\begin{align}
  \label{eq:e0}
  E_0 \equiv F^{+-}_{(0)} &= ig \delta^{ij} \left[ A_1^i,
  A_2^j \right] \, ,\\
  \label{eq:b0}
  B_0 \equiv F^{21}_{(0)} &= ig \epsilon^{ij} \left[ A_1^i,
  A_2^j \right] \, ,
\end{align}
at $\tau=0$. The gauge 
fields $A_{1,2}^i(\vec x_\perp)$, for $i=1,2$, are created by the 
sources $\rho_{1}$ and $\rho_{2}$  in each nucleus before the collision, respectively .

As discussed in detail in Ref.\ \cite{Chen:2013ksa} the decay of the longitudinal
field over time leads to rapidity-even transverse fields at first order
in time through Faraday's and Amp\`ere's Law. On the other hand Gauss' Law 
leads to rapidity-odd components of the transverse fields. In detail, we have
\begin{align}
  E^i_{(1)} &= -\frac{1}{2} \left( \sinh\eta [D^i, E_0] + \cosh\eta \,
  \epsilon^{ij}[D^j,B_0] \right) \, ,
  \label{eq:trfield1}  \\
  B^i_{(1)} &= \frac{1}{2} \left( \cosh\eta\, \epsilon^{ij} [D^j, E_0]
  - \sinh\eta [D^i,B_0] \right) \, ,
  \label{eq:trfield2}
\end{align}
for $i=1,2$, where the transverse covariant derivative is with respect to the field 
on the light cone, $D^i =\partial^i - ig (A^i_1+A^i_2)$, and $\eta=1/2 \ln x^+/x^-$ is the 
space-time rapidity.
The second order in time describes the back reaction of these transverse
fields on the longitudinal fields.
The rapidity-even and -odd transverse fields drive rapidity-even and -odd
terms in the Poynting vector at first order in time,
\begin{align}
  \label{eq:T0i_1}
  T^{0i}_{(1)} =&  
  \frac{1}{2} \alpha^i \cosh\eta  +
     \frac{1}{2} \beta^i \sinh\eta  \, , \\
  \label{eq:T3i_1}
  T^{3i}_{(1)} =& 
  \frac{1}{2} \alpha^i \sinh\eta + \frac{1}{2} \beta^i \cosh\eta \, ,
\end{align}
for $i=1,2$, where the two relevant transverse vectors are \cite{Chen:2013ksa}
\begin{align}
  \alpha^i =& - \nabla^i \varepsilon_0 \, ,\\ 
  \beta^i =& \epsilon^{ij} \left( [D^j,B_0]E_0 - [D^j,E_0]B_0\right) \, .
\end{align}
%The properties of $\alpha^i$, $\beta^i$ will be discussed extensively in the following sections.

In Refs.\ \cite{Chen:2013ksa,Chen:2015wia} the functional integrals over $\rho_1$ and
$\rho_2$ with Gaussian weights were also calculated analytically, resulting 
in closed expressions for event averages. The size of the Gaussian color 
charge fluctuations is fixed by
\begin{equation}
  \label{eq:chargedensnorm}
  \langle \rho_{\underline{a}}(\vec x_\perp) \rho_{\underline{b}}
  (\vec y_\perp) \rangle = \frac{g^2\delta_{\underline{ab}} }{N_c^2-1} \mu(\vec x_\perp) 
  \delta^2 (\vec x_\perp - \vec y_\perp ) \, ,
\end{equation}
where $\mu(\vec x_\perp) $ characterizes the local color charge fluctuation strength,
and the underlined indices are for color degrees of freedom.
The event averaged initial energy density is 
\cite{Lappi:06,Chen:2013ksa,Chen:2015wia} 
\begin{equation}
 \varepsilon_0 (\vec x_\perp) = \left\langle T^{00}_{(0)} \right\rangle = 
\frac{2 \pi N_c \alpha_s^3}{ N_c^2-1 } \mu_1 (\vec x_\perp) \mu_2 (\vec x_\perp)
\ln^2 \left(\frac{Q^2}{\hat m^2}\right) \,,
\label{ini_den}
\end{equation}
where $Q$ and $\hat m$ are ultraviolet and infrared momentum cutoffs respectively,
delimiting the validity of the color condensate model. The $\mu_{1,2}$
are the profiles for color fluctuations in nucleus $1$ and $2$, respectively.

The event averaged contributions to the transverse energy flow are
\begin{align}
  \langle\alpha^i\rangle  &=  -\varepsilon_0 \frac{\nabla^i\left( \mu_1 \mu_2 
  \right)}{2 \mu_1\mu_2}
   \, , \\
  \langle \beta^i \rangle &= - \varepsilon_0 \frac{\mu_2 \nabla^i \mu_1 
  - \mu_1 \nabla^i
  \mu_2}{\mu_1\mu_2}   \, .
\end{align}
Note that these structures are rather universal and also appear in the
energy flow in strong coupling calculations \cite{Romatschke:2013re}. 
We will drop the brackets $\langle\ldots\rangle$ in our notation from here. 
All components of the energy momentum tensor are considered event-averaged 
unless indicated otherwise.

We can summarize the structure of the classical MV energy momentum tensor
as follows. We will write the tensor in Milne metric with coordinates ($\tau,x,y,\eta$). 
Recall that our setup is boost-invariant. In Milne coordinates this translates
into vectors and tensors being translationally invariant in $\eta$, while
in the Cartesian coordinates the $\eta$-dependence is given by longitudinal
Lorentz boosts, leading to the familiar $\cosh\eta$ and $\sinh\eta$ factors.
In terms of the Milne components we have
\begin{widetext}
\begin{equation}
  T^{m n}_{\mathrm{YM}} =
  \begin{pmatrix}
    \varepsilon_0-\frac{\tau^2}{8}(-2\triangle \epsilon_0 + \delta)  & \frac{\tau}{2} \alpha^1 + \frac{\tau^3}{16} \xi^1 & \frac{\tau}{2} \alpha^2 + \frac{\tau^3}{16} \xi^2  & -\frac{\tau}{8} \nabla^i \beta^i \\
    \frac{\tau}{2} \alpha^1 + \frac{\tau^3}{16} \xi^1  & \varepsilon_0-\frac{\tau^2}{4}(-\triangle \varepsilon_0 + \delta-\omega) & \frac{\tau^2}{4}\gamma & \frac{1}{2} \beta^1+ \frac{\tau^2}{16} \zeta^1  \\
    \frac{\tau}{2} \alpha^2 + \frac{\tau^3}{16} \xi^2  & \frac{\tau^2}{4}\gamma & \varepsilon_0-\frac{\tau^2}{4}(-\triangle \varepsilon_0 + \delta+\omega) & \frac{1}{2} \beta^2 + \frac{\tau^2}{16} \zeta^2  \\
    -\frac{\tau}{8} \nabla^i \beta^i    & \frac{1}{2} \beta^1 + \frac{\tau^2}{16} \zeta^1 & \frac{1}{2} \beta^2 + \frac{\tau^2}{16} \zeta^2 & -\frac{\varepsilon_0}{\tau^2} + \frac{1}{8}(-2\triangle \varepsilon_0 + 3 \delta)
      \end{pmatrix} + \mathcal{O}(\tau^4)  \, ,
      \label{eq:tmn}
\end{equation}
\end{widetext}
where explicit expressions for the higher order quantities $\delta$, $\omega$, 
$\gamma$, $\xi^i$ and $\zeta^i$ ($i=1,2$) can be found in Ref.\ \cite{Chen:2015wia}.

\subsection{Angular Momentum of the Gluon Field}
\label{sec:ymang}

The covariant angular momentum density of a relativistic system with energy momentum tensor
$T^{\mu\nu}$ is given by the rank-3 tensor
\begin{equation}
  \label{eq:mtensor}
  M^{\mu \nu \lambda}= r^\mu T^{\nu \lambda} - r^\nu T^{\mu \lambda}  \, ,
\end{equation}
where $r^\mu$ is the position with respect to a reference point in Minkowski
space. For our purposes we will choose the reference point as the center
of the usual coordinate system. We have already fixed $t=0$, $z=0$ at 
nuclear overlap, and we choose $x=0$, $y=0$ to be the point at the center,
i.e.\ halfway along the impact vector in the transverse plane. The impact vector
also determines the direction of the $x$-axis. For event-averaged 
collisions there is no problem with fluctuations and the event plane
is readily defined as the $x$-$z$-plane.

Let us explore the angular momentum contents of the classical Yang-Mills
field at early times given by the energy momentum tensor in Eq.\ (\ref{eq:tmn}).
With the coordinate parameterized as $r^\mu =(\tau\cosh\eta,x,y,\tau\sinh\eta)$
and restricting ourselves to order $\mathcal{O}(\tau^2)$ in 
the energy momentum tensor we find
\begin{align}
  M_{\mathrm{YM}}^{120} =& \frac{\tau}{2}\cosh\eta \left( x \alpha^2 -y
  \alpha^1 \right)   \nonumber
  \\ &+ \frac{\tau}{2} \sinh\eta \left( x \beta^2 -y \beta^1
  \right)  + \mathcal{O}(\tau^3) \label{eq:l1} \, ,  
  \\
  M_{\mathrm{YM}}^{310} = & \frac{\tau^2}{2}\sinh\eta
  \left( \cosh\eta \alpha^1+\sinh\eta \beta^1\right)
  \\ & -\frac{\tau^2 x}{8}\left( \delta \sinh 2\eta - \nabla^i\beta^i \cosh 2\eta
  \right) +\mathcal{O}(\tau^4)  \, ,  \nonumber 
  \\
  M_{\mathrm{YM}}^{230} = & -\frac{\tau^2}{2}\sinh\eta
  \left( \cosh\eta \alpha^2+\sinh\eta \beta^2\right)  \label{eq:l3}
  \\ &
  +\frac{\tau^2 y}{8}\left( \delta \sinh 2\eta - \nabla^i\beta^i \cosh 2\eta
  \right) +\mathcal{O}(\tau^4)  \, . \nonumber 
\end{align}
Most of the 64 components of the angular momentum tensor do not vanish and
some more will be discussed below. For now we focus on these three specific 
components since they are related to the usual angular momentum vector $\mathbf{L}=(L_1,L_2,L_3)$ 
in a volume $V$ as
\begin{equation}
  L_i = \frac{1}{2} \epsilon^{ijk} \int_V d^3 r M^{jk0}   \, ,
\end{equation}
for $i=1,2,3$.

In a boost-invariant system the notion of angular momentum only makes sense 
per unit of space-time rapidity. The total angular momentum cannot be expected
to be finite or globally conserved as the fixed charges on the light cone
act as sources of energy, momentum and angular momentum. Nevertheless in
regions of the collision for which boost-invariance is a good approximation 
we expect the angular momentum per rapidity $\dif \mathbf{L}/{\dif \eta}$ to be 
a meaningful quantity. Let us begin by considering
\begin{equation}
  \frac{\dif L_3}{\dif \eta} = \tau\int \dif^2 r_\perp M_{\mathrm{YM}}^{120} \, 
\end{equation}
where $\vec r_\perp = (x,y)$ is the transverse coordinate vector.
In a collision of equal spherical nuclei A+A at any impact parameter, 
or for an asymmetric collision A+B of spherical nuclei at zero impact 
parameter, we can use the apparent symmetries in the transverse plane to argue 
that all terms appearing in the transverse integration of 
the angular momentum density $M_{\mathrm{YM}}^{120}$ vanish. 
We focus on such collisions for the rest of this subsection, noting that
Au+Au collisions and Pb+Pb collisions at any impact parameter will be covered.
%The symmetry arguments we use are valid for all event-averaged collisions of
%that type.

\begin{table}[tb]
\begin{center}
\begin{tabular}{|c|c|c|}
  \hline
  & $x$ & $y$ \\ \hline
  $\varepsilon_0$ & even & even \\ \hline
  $\alpha^1$ & odd & even \\ \hline
  $\alpha^2$ & even & odd \\ \hline
  $\beta^1$ & even & even \\ \hline
  $\beta^2$ & odd & odd \\ \hline
  $\nabla^i \alpha^i$ & even & even \\ \hline
  $\nabla^i \beta^i$ & odd & even \\ \hline
  $\delta$ & even & even \\ \hline
  $\omega$ & even & even \\ \hline
  $\gamma$ & odd & odd \\ \hline
\end{tabular}
\caption{Parity with respect to the $x$ and $y$ coordinate of all 
coefficients appearing in the power series of the Yang-Mills energy 
momentum tensor up to second order in $\tau$ \cite{Chen:2015wia}. The results apply to
event averaged symmetric A+A collisions and event averaged A+B collisions
at vanishing impact parameter. 
\label{tab:parity}}
\end{center}
\end{table}

For example, from the explicit expressions in Ref.\
\cite{Chen:2015wia} we find that for A+A collisions 
\begin{multline}
  \epsilon_0(x,y) \sim \mu_1(x,y)\mu_2(x,y) \\
  = \mu\left( x-\frac{b}{2},y\right)
  \mu\left( x+\frac{b}{2},y\right) \, ,
\end{multline}
is even as a function of both $x$ and $y$. For A+B collisions at finite
impact parameter the last identity does not hold but the conclusion about
parity with respect to both coordinates is valid nevertheless. It follows 
that, e.g., $\alpha^1 = - \partial\varepsilon_0/\partial x $ is odd in 
coordinate $x$ and even in coordinate $y$. The symmetries of all relevant 
terms with respect to both coordinates are summarized in Tab.\ \ref{tab:parity}.
We can now readily determine that terms like $y \alpha^1$ and $x\alpha^2$
are odd in $y$ and vanish when integrated over the transverse plane.
In fact most terms in Eqs.\ (\ref{eq:l1})-(\ref{eq:l3}) disappear upon
integration over transverse coordinates, and we arrive at the expressions
\begin{align}
  \frac{\dif L_2}{\dif \eta} = & \, \frac{\tau^3}{2} \int \dif^2 r_\perp
  \left( \sinh^2\eta \,\beta^1 + \cosh 2\eta\, \frac{x}{4} \nabla^i\beta^i 
  \right) \, ,  \nonumber
  \\
  \frac{\dif L_1}{\dif \eta} =& \, 0 = \frac{\dif L_3}{\dif \eta}
  \, 
  \label{eq:angmom1}
\end{align}
for the early time gluon field, up to third order in time.

This result is not surprising. First, the system of 
colliding nuclei, for impact parameter $b\ne 0$, carries total orbital angular momentum 
$L_2^0 \ne 0$ (in our choice of coordinate system), while $L_1^0 =0=L_3^0$. The 
primordial angular momentum $L_2^0$ in $y$-direction leaves an imprint on the 
gluon field created after the collision. Moreover, the rapidity-odd flow field 
($\beta^i$ at the lowest order in $\tau$) is the carrier of angular momentum in the gluon
field.
To be more precise, there are two contributions to the local angular momentum $dL_2/\eta$ in 
$y$-direction, both symmetric in rapidity. The first one, vanishing at $\eta=0$ is the
directed flow $\beta^i$ itself. The second one is also present at midrapidity
and comes from a rapidity-even term in $T^{03}$, i.e.\ the longitudinal flow of energy.
It can be identified with a longitudinal shear flow: the longitudinal flow
of energy of the gluon field moves in opposing directions for $x>0$ and
$x<0$. In terms of the dynamical evolution it is a response to the build up
of directed flow, but both terms contribute at the same cubic power of time
to the angular momentum. The terms are more closely related than it
appears at first sight and we simplify the expression for $dL_2/d\eta$ 
further below.

%Before we do that let us briefly analyze other components of the angular 
%momentum tensor.
%For completeness we mention the dynamical mass vector 
%\begin{equation}
%  K_i = - \int_V d^3r M^{0i0}_{\mathrm{YM}}  \, ,
%\end{equation}
%for $i=1,2,3$,
%which generates boosts and mixes with angular momentum in relativistic Lorentz
%boosts. Again integrated over the transverse plane for symmetric A+A
%collisions, or for vanishing impact parameter collisions, we obtain
%\begin{align}
 % \frac{\dif K_1}{\dif \eta} =& \,  - \frac{\tau^3}{4}\sinh 2\eta \int \dif^2r_\perp 
  %\left( \beta^1 + \frac{1}{2} x\nabla^i\beta^i \right)  \, , \\
 % \frac{\dif K_2}{\dif \eta} = & \, 0 \, ,  \\
%  \frac{\dif K_3}{\dif \eta} =& \, \tau^2 \sinh \eta \int   \dif^2r_\perp \varepsilon_0
%  \, ,
%\end{align}
%up to the third order in time.
%More interesting for our purposes are 
Let us investigate the components of the angular momentum 
tensor that encode the \emph{flow} of angular momentum $L_2$, namely
\begin{align}
  M^{311}_{\mathrm{YM}} =& \, \tau\sinh\eta\, \varepsilon_0 \\ &- \frac{\tau x}{2}
  \left(\sinh\eta\,  \alpha^1 +\cosh\eta\, \beta^1\right)
   + \mathcal{O}(\tau^2) \,
  ,  \nonumber \\
  M^{312}_{\mathrm{YM}} =& - \frac{\tau x}{2}
  \left(\sinh\eta\,  \alpha^2 +\cosh\eta\, \beta^2\right)
  + \mathcal{O}(\tau^2)\, ,
  \\
  M^{313}_{\mathrm{YM}} =& \frac{\tau^2}{2}\sinh\eta\left( \sinh\eta\,  \alpha^1 
  +\cosh\eta \, \beta^1\right) \nonumber \\ & - x \left( -\varepsilon_0 +\frac{\tau^2}{4}
  \left( \nabla^i\alpha^i + \delta\right)\right.  \\
  & \left. -\frac{\tau^2}{8} \left( \sinh 2\eta\, 
  \nabla^i\beta^i - \cosh 2\eta\, \delta \right)\right) + \mathcal{O}(\tau^3) 
  \, . \nonumber
\end{align}
One can verify readily that $L_2$ is a conserved quantity, as expected,
by working out that $\partial_\mu M^{31 \mu}_{\mathrm{YM}} = 0$ up to first order in time.

Returning to an analysis of the angular momentum per slice in rapidity
we note that the \emph{transverse} flow of angular momentum should disappear in that
case. Indeed we find
\begin{multline}
  0 = \int \dif^2 r_\perp \nabla^i M^{31i}_{\mathrm{YM}} \\ 
   = -\frac{\tau}{2} \cosh\eta \int \dif^2 r_\perp
  \left( \beta^1+ x \nabla^i\beta^i \right) \; ,
  \label{eq:notrflow}
\end{multline}
where we sum over transverse indices $i=1,2$. The first identity holds
according to Gauss' Theorem for any reasonably fast falling functions under
the integral,
%$M^{31i}_{\mathrm{YM}}$ 
which in our case is enforced by the nuclear profile functions 
$\mu_{1,2}$ going to zero rapidly outside the nuclei. 
This leads to the relation between the integrals over $\beta^1$ and
$x \nabla^i\beta^i$ alluded to before. We can use this identity to arrive
at the final versions for the angular momentum per rapidity and its
longitudinal flow $N_2 =\frac{1}{2} \int_V d^3 r M^{313}$. We obtain
\begin{align}
    \frac{\dif L_2}{\dif \eta} = & \, \frac{\tau^3}{8}\left(\sinh^2\eta-1\right)
   \int \dif^2 r_\perp \beta^1 \, , \\
  \frac{\dif N_2}{\dif \eta} = & \, \frac{\tau^3}{8} \sinh 2\eta
   \int \dif^2 r_\perp \beta^1
   \, .
\end{align}
According to Eq.\ (\ref{eq:notrflow}) the conservation law for angular 
momentum flow between rapidity slices reduces to
\begin{equation}
  \frac{\partial}{\partial t} \frac{\dif L_2}{\tau \dif \eta} + 
 \frac{\partial}{\partial z} \frac{\dif N_2}{\tau \dif \eta} =0 \; ,
\end{equation}
which can be checked explicitly.

\begin{figure}[tb]
\includegraphics[width=\columnwidth]{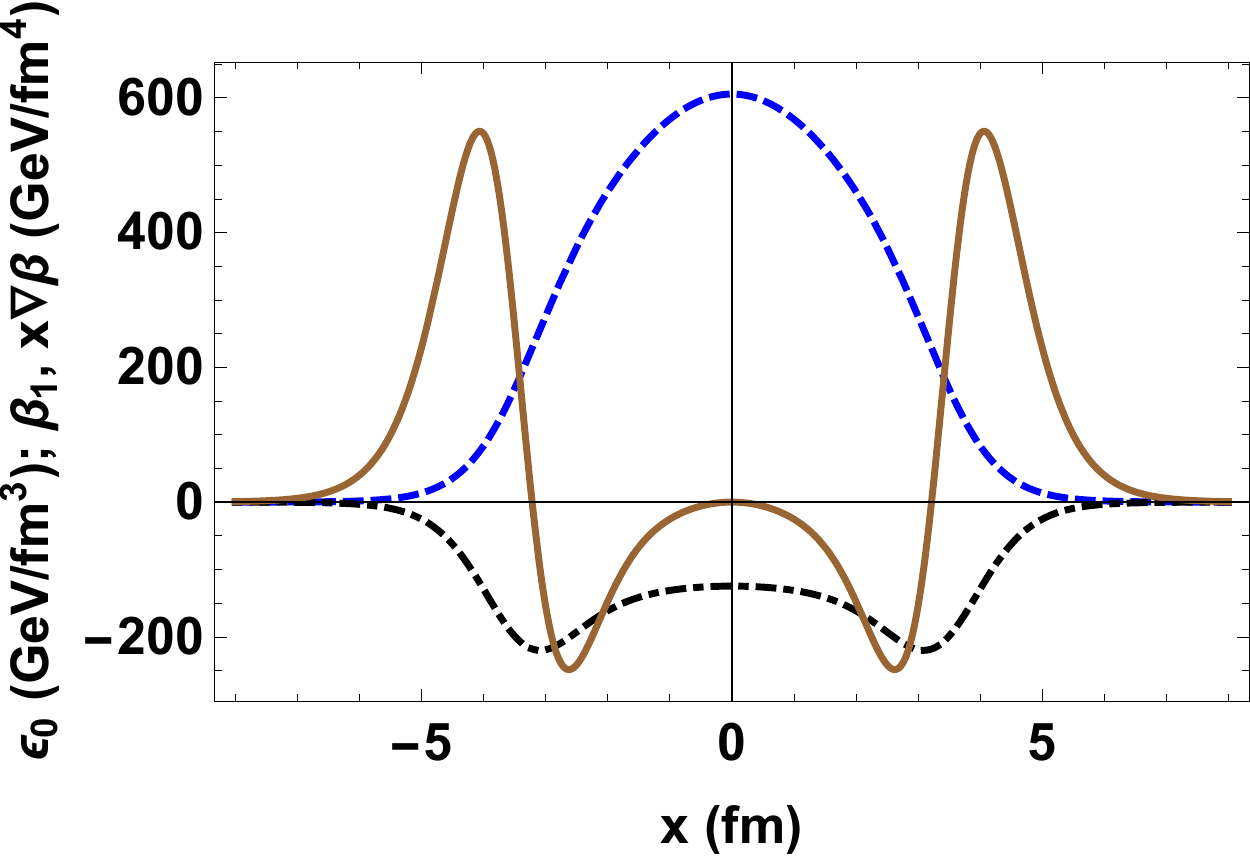}
\caption{\label{fig:delbeta} The initial energy density $\varepsilon_0$ 
 (dashed), directed flow in $x$-direction $\beta^1$ (dash-dotted), 
 and the divergence of directed flow $x\nabla^i\beta^i$ weighted by $x$ (solid line) 
 as functions of coordinate $x$ at $y=0$ for
 Pb+Pb collisions at impact parameter $b=6$ fm.}
\end{figure}

Let us now explore these results numerically. We consider as an example
Pb+Pb collisions. We model the shape of the charge fluctuation densities 
$\mu_{1,2}(x,y)$ by the nuclear thickness functions of the nuclei as given 
by Woods-Saxon distributions with appropriate radius $R_A$.
We take nucleus $1$ to be traveling in the $+z$-direction with center at
$x=b/2$ and nucleus $2$ to be traveling in the $-z$-direction with center
at $x=-b/2$. Thus the primordial total angular momentum of the system
at a finite total center of mass energy $\sqrt{s}$ is
\begin{equation}
  L_2^0 = - \frac{b}{2} \sqrt{s} \, ,
\end{equation}
and $L_1^0=0=L_3^0$. The primordial angular momentum is clockwise
in the reaction plane, $L_2^0 < 0$.

The important quantities for the initial angular momentum after the collision
are $\beta^1$ and $x \nabla^i\beta^i$. They are
shown as functions of $x$ for collisions with impact parameter $b=6$ fm
in Fig.\ \ref{fig:delbeta}. We note that $\beta^1$ is negative, i.e.\ for
$\eta>0$ the directed flow is in the negative $x$-direction as expected for
directed flow, and its contribution to the angular momentum $dL_2/d\eta$ 
is negative. We already know from Eq.\ (\ref{eq:notrflow}) that the contribution from shear flow will
have the opposite sign. That seems strange since that flow is opposing
the motion of the initial nuclei. However,
Fig.\ \ref{fig:delbeta} makes it clear that $x\nabla^i\beta^i$ must have two nodes 
along the $x$-axis toward the outer regions of the fireball where its sign 
flips. In the inner regions the sign is indeed negative, leading to clockwise
shear flow, while the outer regions exhibit a counter-clockwise shear flow.
Upon integration over the transverse plane the counter-clockwise shear flow
wins due to the weighting with the lever arm $x$.

\begin{figure}[tb]
\includegraphics[width=\columnwidth]{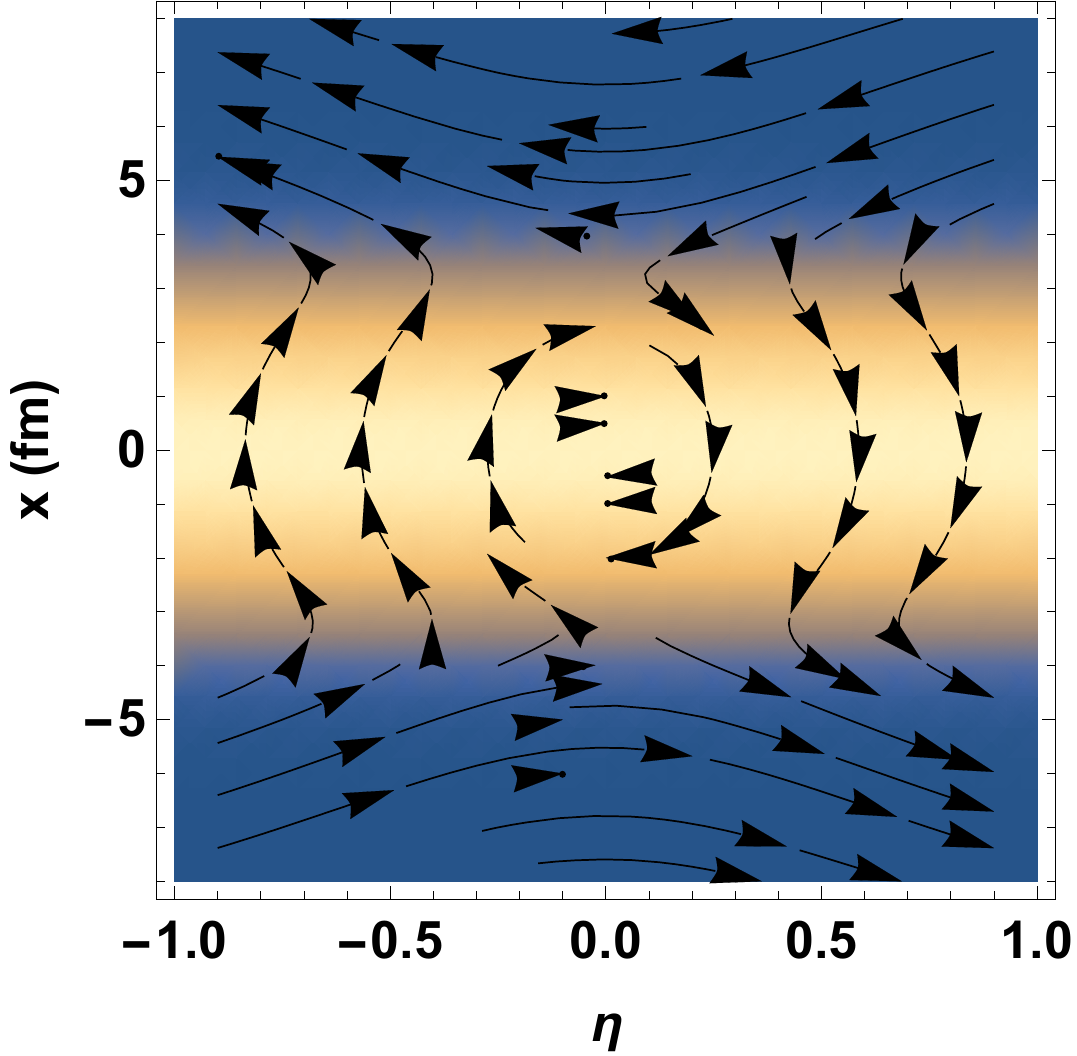}
\caption{\label{fig:oddflow} Energy flow components $(T^{01},T^{03})_{\mathrm{odd}} 
\sim (\tau\beta^1 , -\tau^2\nabla^i\beta^i /4)$ contributing to local angular momentum $dL_2/d\eta$, up 
to second order in time, plotted at $\tau=0.25$ fm/$c$ with the initial energy density 
 $\varepsilon_0$ in the background, for Pb+Pb collisions at impact parameter $b=6$ fm.}
\end{figure}

\begin{figure}[tb]
\includegraphics[width=\columnwidth]{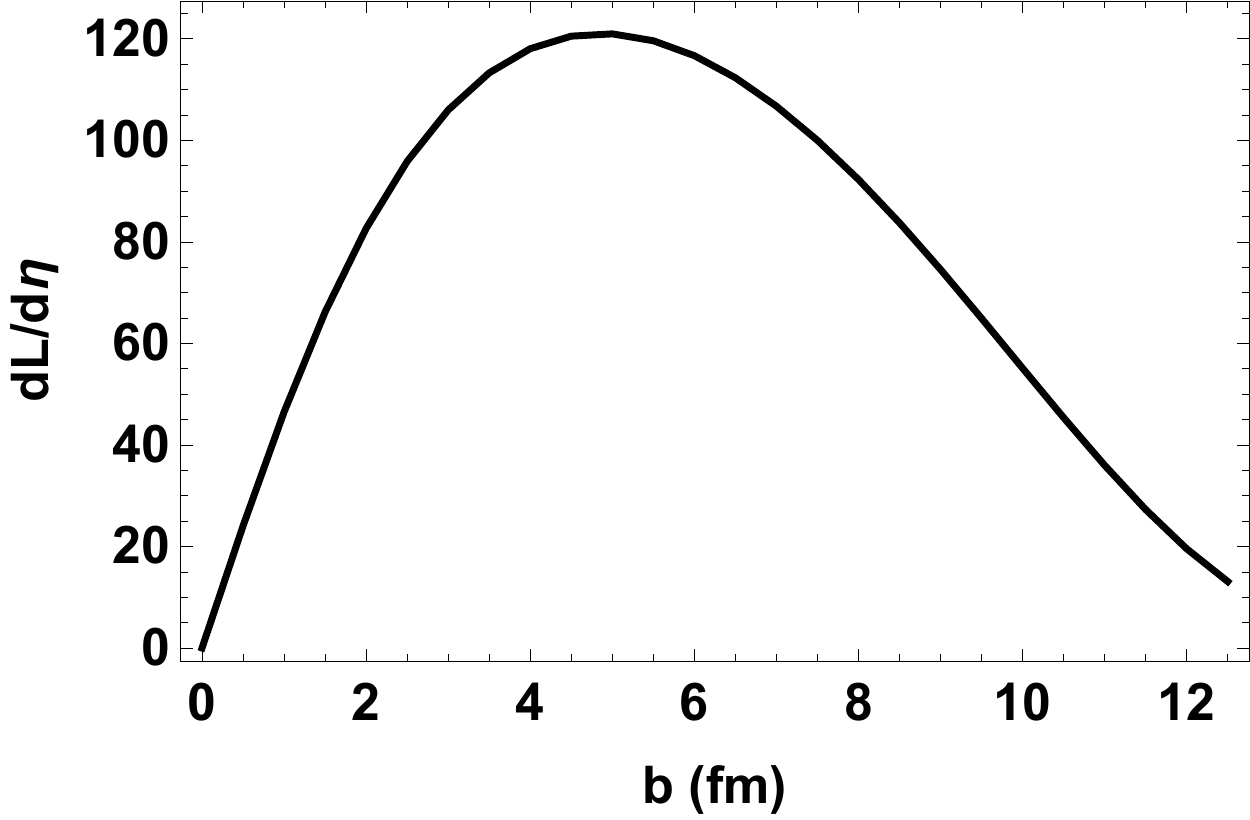}
\caption{\label{fig:Lvsb} The angular momentum per rapidity $dL_2/d\eta$
  at $\eta=0$ at a time $\tau=0.25$ fm/$c$ as a function of impact parameter $b$ for Pb+Pb 
  collisions as given by the leading terms in the power series in time.}
\end{figure}

An interesting picture emerges. Fig.\ \ref{fig:oddflow} shows
the part of the energy flow vector $(T^{01},T^{03})$ contributing to angular momentum
in the reaction plane. To be precise, we plot the rapidity-even part of the 03-component and the 
rapidity-odd part of the 01-component, 
$(T^{01},T^{03})_{\text{odd}} = (T^{01}(\eta)-T^{01}(-\eta),T^{03}(\eta) + T^{03}(-\eta))/2$.
This procedure drops terms proportional to $\alpha^1$ or $\nabla^i\alpha^i$ which act as a background here.
We can clearly recognize the clockwise eddy current of energy flow at the center, and the
counter-clockwise shear flow in the outer regions. Note that the nucleus in
the top half ($x>0$) is moving to the right, the one at the bottom ($x<0$) is 
moving to the left.  

In Fig.\ \ref{fig:Lvsb} we show the angular momentum
$\dif L_2/\dif \eta$, at $\eta=0$, as a function of the impact parameter $b$ for Pb+Pb 
collisions. As expected the angular momentum of the gluon field goes to 
zero for $b=0$ and for $b \gg R_A$. Recall that this result holds for event averages
with color fluctuation densities $\mu_{1,2}$, assumed to follow Woods-Saxon distributions.
The shape of the curve is qualitatively consistent with the result from \cite{Becattini:2007sr} in a 
setup inspired by the Glauber model.
The numerical value of the angular momentum per rapidity is determined by
the normalization of the initial energy density $\varepsilon_0$ and depends on
time. For the former we have chosen $\varepsilon_0 = 605$ GeV/fm$^3$ at a time $\tau=0.2$ fm/$c$ 
at the center for $b=6$ fm. This value has the correct order of magnitude for collisions at LHC energies
but has not been tuned to data with any precision.
One can estimate the relation between $\dif L_2/\dif \eta$ and the initial energy density $\varepsilon_0$ when
one uses a simple approximation for the profile functions $\mu_{1,2}(x,y)$. Let
us assume the profile functions represent simple slabs of radius $R_A$, with a fixed value $\bar\mu$
inside the slab. Then $\varepsilon_0$ is also a function describing a slab in the
transverse plane with a value of $\bar\varepsilon_0 \sim \bar\mu^2$ inside the overlap region
of the nuclei. From Ref.\ \cite{Chen:2015wia} we find
\begin{multline}
  \beta^1 = \varepsilon_0  \frac{\mu_1 \nabla^1\mu_2-\mu_2\nabla^1\mu_1}{
  \bar\mu^2}  \\
  = \varepsilon_0 \left( -\delta(x-\partial_2) - \delta(x-\partial_1) \right) \; ,
\end{multline}
where $\partial_i$ here denotes the $x$-coordinate of the boundary of slab $i$
(at a fixed $y$) within the other slab. Note that the last expression means
that we have approximated the realistic function $\beta^1(x)$ shown in Fig.\ 
\ref{fig:delbeta} with two negative $\delta$-functions that replace the 
two negative peaks. We can easily integrate over $x$ and approximate the
size of the fireball to be roughly $2R_A$ in $y$-direction. Thus we
obtain approximately
\begin{equation}
  \int \dif^2r_\perp \beta^1 \approx -4 \bar \varepsilon_0 R_A \; ,
\end{equation}
for $0<b<2R_A$, and we arrive at the final estimate
\begin{equation}
  \label{eq:l2est}
  \frac{\dif L_2}{\dif \eta} \approx \frac{1}{2} \tau^3 R_A \bar\varepsilon_0  
  \left( 1-\sinh^2\eta\right)  \; .
\end{equation}
This result should still hold approximately if we relax the assumption
of slabs and treat $\bar\varepsilon_0$ as an average initial energy density of
a system with more realistic transverse profiles. Let us emphasize once more, that
despite the vortex aligned with the primordial total angular momentum, if integrated
over the entire transverse plane $dL_2/d\eta >0$ is opposite to $L_2^0$. It is created by
an \emph{ingoing} longitudinal angular momentum flux, 
$dN_2/d\eta <0$ for $\eta>0$.

Lastly, let us note that in single events which deviate from the situation discussed
here by fluctuations, could exhibit interesting additional local dynamics. Such calculations
would have to be carried out numerically. However the global picture that we have 
presented here by integrating over the transverse plane should stay intact 
when transverse fluctuations are introduced.

\section{\label{sec:match} Matching to Fluid Dynamics}

The energy momentum tensor of a system which is locally in full kinetic 
equilibrium can be written as
\begin{equation}
 \label{eq:id}
 T_{\mathrm{id}}^{\mu\nu}= (e + p) u^\mu u^\nu - p g^{\mu \nu}  \, ,
\end{equation}
where $e$ and $p$ are the local energy density and (equilibrium) pressure,
related by the equation of state, and $u^\mu$ is the flow velocity of each local
fluid cell as seen by the observer.
Deviations from local kinetic equilibrium make it necessary to introduce
dissipative corrections in the form of bulk stress $\Pi$ and shear stress tensor 
$\pi^{\mu\nu}$. The energy momentum tensor of viscous fluid dynamics
can then be expressed as \cite{Muronga:2004sf}
\begin{equation}
  \label{eq:vf}
 T_{\mathrm{vf}}^{\mu\nu}= (e + p +\Pi) u^\mu u^\nu - (p+\Pi) g^{\mu \nu} + \pi^{\mu \nu} \; ,
\end{equation}
where the tensor $\pi^{\mu\nu}$ is symmetric, traceless and orthogonal to the 
flow velocity, $u_\mu \pi^{\mu\nu} = 0$. We have chosen to neglect additional
conserved currents, such as baryon number, as we aim to match
to a pure Yang-Mills system in this work. We have also lifted the ambiguity
coming from the incomplete definition of a local rest frame by setting the
heat flow $q^\mu=0$ to zero (Landau frame).
%It is an interesting question is any symmetric energy momentum tensor can be decomposed
%as in (\ref{eq:vf}).

\subsection{Matching to Ideal Fluid Dynamics}

We will first discuss the case of a system undergoing rapid thermalization
at a time $\tau_{\mathrm{th}}$,
so that at the end of a rather short time period the energy momentum 
tensor can be written as in Eq.\ (\ref{eq:id}). 
It is assumed that during the short time interval in which local equilibrium is achieved, due to microscopic processes,
the macroscopic evolution given by the flow of energy and momentum is negligible.
In this instantaneous approximation we can hope to write the total energy
momentum tensor as 
\begin{equation}
  T^{\mu\nu}_{\mathrm{tot}}=\Theta(\tau_{\mathrm{th}}-\tau)T^{\mu\nu}  +\Theta(\tau-\tau_{\mathrm{th}})T_{\mathrm{id}}^{\mu\nu} \; ,
\end{equation} 
where we have denoted the tensor before equilibration simply as $T^{\mu\nu}$, 
and $\Theta$ is the Heaviside step function.

Clearly not all energy momentum tensors can be written as an 
equilibrium tensor. Negative pressure, for example realized
in Yang-Mills fields discussed here, see Eq.\ (\ref{eq:tmn}),
cannot be accommodated in kinetic equilibrium. Thus the components of
the energy momentum tensor must be permitted to change rapidly during 
the short equilibration phase.
In order to constrain the local energy density $e$ and the 3-velocity 
$\mathbf{v}$, with $u^\mu=\gamma(1,\mathbf{v})$ at the end of this 
rapid equilibration process, we can use the fact that we have to
satisfy energy and momentum conservation. Given the hierarchy of time scales
in our assumptions we impose \cite{Fries:2005yc,Fries:2007iy}
\begin{equation}
  \label{eq:epmatch}
  \partial_\mu T^{\mu\nu}_{\mathrm{tot}}=0 \, .
\end{equation}
This condition provides four equations which equals the number of unknowns
($e, \mathbf{v})$. The pressure needs to be determined self-consistently
in the equilibrated phase by the equation of state $p=p(e)$.
Of course $\partial_\mu T^{\mu\nu}=0=\partial_\mu T^{\mu\nu}_{\mathrm{id}}$
and thus Eq.\ (\ref{eq:epmatch}), works out to the simple condition
\begin{equation}
  n_\mu T^{\mu \nu} = n_\mu T^{\mu\nu}_{\mathrm{id}} \, ,
\end{equation}
for $\nu=0,1,2,3$. Here $n_\mu = (\cosh \eta,0,0,-\sinh\eta)$ is the normal
vector of the matching hypersurface $\tau=\tau_{\mathrm{th}}$.
Hence, though energy and momentum are conserved at every point of the matching 
hypersurface, only the projection of $T^{\mu\nu}_{\mathrm{tot}}$ perpendicular
to the matching surface is forced to be continuous at $\tau=\tau_{\mathrm{th}}$.
Indeed, the longitudinal pressure in the case of the Yang-Mills energy 
momentum tensor from the last section matched to ideal fluid dynamics 
provides an example for non-continuous components.

In the case of a boost-invariant Yang-Mills energy momentum tensor 
as in the previous section, we can use the boost-symmetry and restrict
ourselves to determining the fluid dynamic fields at the space-time rapidity 
slice $\eta=0$ only. Lorentz boosts then 
provide the result at any other $\eta$. We find a set of four equations   
\begin{align}
   e =& \, \varepsilon - \frac{S^2}{\varepsilon+p}  \; ,
  \nonumber \\
  \mathbf{ v} =& \, \frac{\mathbf{S}}{\varepsilon+p}    \; ,
  \label{eq:matchideal1}
\end{align}
where $\varepsilon=T^{00}$ and $S^i=T^{0i}$, for $i=1,2,3$, are the energy density and
Poynting vector of the Yang-Mills system at midrapidity, respectively. 
Recall that the equation of state for the equilibrium pressure $p$ closes 
the system of equations.

Using the expressions up to second order in time from Eq.\ (\ref{eq:tmn}) 
the equations can be cast in the form
\cite{Fries:2005yc,Fries:2007iy}
\begin{align}
   e =& \, \varepsilon - \frac{\tau_{\mathrm{th}}^2}{4}
  \frac{\alpha^2+\frac{\tau_{\mathrm{th}}^2}{16}(\nabla^k \beta^k)^2}{
  \varepsilon+p}  \; ,\nonumber \\
  v_i =& \, \frac{\tau_{\mathrm{th}}}{2} \frac{\alpha^i}{\varepsilon+p} \, ,
    \> i=1,2\, ,  \label{eq:matchideal2} \\
  v_3 =& \, - \frac{\tau_{\mathrm{th}}^2}{8}\frac{\nabla^k \beta^k}{\varepsilon+p} \; ,
  \nonumber
\end{align}
$i=1,2$, with
\begin{equation}
  \varepsilon = \varepsilon_0 + \frac{\tau_{\mathrm{th}}^2}{8}\left(2\triangle
  \epsilon_0 - \delta\right) \, . 
 \end{equation}

First we notice that the rapidity-odd flow $\beta^i$ does not directly enter
the expression for the transverse flow velocity $v_i$, for $i=1,2$. In fact the 
direction of the transverse fluid flow field $v_i$ is entirely 
determined by the direction of the Poynting vector at midrapidity, 
which is given by the rapidity-even flow $\alpha^i$. Information about 
directed flow is lost in the matching procedure, consistent with the
possibility that individual components of the energy momentum tensor can 
be discontinuous on the matching
hypersurface. On the other hand, the longitudinal velocity flow $v_3$ is 
directly proportional to the longitudinal gluon energy flow term 
$-\nabla^i\beta^i$. Therefore longitudinal shear flow is introduced into 
the fluid velocity field.

Turning to the angular momentum we note that $\partial_\mu T^{\mu\nu}=0$ for
any \emph{symmetric} energy momentum tensor automatically guarantees
$\partial_\lambda M^{\mu\nu\lambda}=0$. So angular momentum is conserved at
the matching surface. However, once more only the projection perpendicular to the matching hypersurface, 
$M^{\mu\nu\lambda} n_\lambda$, has to be continuous across
$\tau=\tau_{\mathrm{th}}$. This ensures that the $\tau$-component of the $L_2$-flow
in Milne coordinates,  
\begin{equation}
  \frac{\dif H_2}{\dif \eta}=  
  \cos\eta \frac{\dif L_2}{\dif \eta} - \sinh\eta \frac{\dif N_2}{\dif \eta} \, 
\end{equation}
is smooth at matching.
However, except at $\eta=0$, $\dif L_2/\dif \eta$ itself is not necessarily continuous across
$\tau=\tau_{\mathrm{th}}$. It is straightforward to test
the smoothness of $\dif H_2/\dif \eta$ at midrapidity explicitly using 
Eq.\ (\ref{eq:matchideal2}) in the calculation of $M^{\mu\nu\lambda}_{\mathrm{id}}$.

The matching to ideal fluid dynamics might seem academic, however it provides
some features that generalize to viscous fluid dynamics. Moreover, an argument could be made
that it is more physical than the popular procedure of matching to viscous fluid dynamics and then
dropping dissipative stress terms.
For the sake of studying the time evolution of angular momentum it will be preferable
to match to viscous fluid dynamics while keeping dissipative stress, as discussed below.

\subsection{Matching To Viscous Fluid Dynamics}

The premise for matching to viscous fluid dynamics is different from the ideal case.
The decomposition in Eq.\ (\ref{eq:vf}) has in principle sufficient degrees of 
freedom and the Yang-Mills energy momentum tensor could be written
directly as the energy momentum tensor of an off-equilibrium fluid. In other
words one would directly seek a solution of the system of equations
\begin{equation}
  T^{\mu\nu}_{\mathrm{YM}} = T^{\mu\nu}_{\mathrm{vf}} \; ,
  \label{eq:vmatch}
\end{equation}
for the 10 independent unknown  fields $e$, $p+\Pi$, $u^\mu$, $\pi^{\mu\nu}$.
The separate determination of $p$ and $\Pi$ can be finalized after choosing an
equation of state on the viscous fluid side. In this paradigm it is obvious that all
components of the energy momentum tensor and the angular momentum tensor 
are continuous, and energy, momentum and angular momentum are trivially
conserved at the matching time. However, to ensure these conservation laws the
viscous fluid dynamics must be initialized with the dissipative stress obtained from the
classical Yang-Mills phase. 

The further approach of the system to equilibrium can then be calculated
with suitable equations of motion for the viscous fluid. This could be standard 
Israel-Steward viscous fluid dynamics \cite{Israel:1979wp,Heinz:2005bw} or 
anisotropic fluid dynamics \cite{Martinez:2010sc,Bazow:2013ifa}. The
latter might be more reliable with the large dissipative corrections we expect 
from a classical Yang-Mills system. In this work we focus on standard viscous
fluid dynamics. We note that although the energy momentum tensor is continuous across the
matching hypersurface, some of the equations of motion are not continuous. 

In general, equation (\ref{eq:vmatch}) might not allow a physically
acceptable unique solution for the fluid fields.
Acceptable here means that the physical conditions, $e>0$, $u^2=1$ with 
$u^0>0$, as well as $\pi^{\mu\nu} u_\nu = 0$ are met. It is useful for practical purposes to 
use the eigenvalue property of the local rest frame
\begin{equation}
  T^\mu_{\,\,\nu} u^\nu = e u^\mu  \, ,
  \label{eq:eval}
\end{equation}
and to state that Eq.\ (\ref{eq:vmatch}) has a unique acceptable solution if
Eq.\ (\ref{eq:eval}) has a unique acceptable solution.
We will solve Eq.\ (\ref{eq:vmatch}) by first solving the eigenvalue problem Eq.\ (\ref{eq:eval}).
which has become standard practice for matching with fluid dynamics, even if the dissipative parts of the
energy momentum tensor are later discarded. 

We can proceed and determine the dissipative stress from
\begin{equation}
  (p+\Pi) \left(u^\mu u^\nu - g^{\mu\nu} \right) + \pi^{\mu\nu} = 
  T^{\mu\nu}_{\mathrm{YM}} - e u^\mu u^\nu  \, ,
  \label{eq:pressure}
\end{equation}
and the equation of state $p=p(e)$.
We can further decompose the contributions to the left hand side by taking the trace,
\begin{equation}
  p+\Pi = \frac{1}{3} \left( e - T^\mu_{\mathrm{YM},\mu}  \right)  \, .
  \label{eq:ppi}
\end{equation}
In the classical Yang-Mills case the initial energy momentum tensor is conformal and thus the 
bulk stress is fixed to one third of the interaction measure $e-3p$ of the fluid after matching.

%The problem of the existence and uniqueness of an acceptable solution can be studied
%for Eq.\ (\ref{eq:eval}) instead of the full problem. For Yang-Mills energy momentum tensors one can
%find explicit counter examples for both uniqueness and existence. However the 
%interesting question arises if these are ``pathological'' cases that can be
%ignored for our practical purposes here. Interestingly, 
%solving Eq.\ (\ref{eq:eval}) numerically for many relevant examples of the tensor in Eq.\
%(\ref{eq:tmn}) have always yielded a solution. Thus we will reserve a further
%discussion of this issue for elsewhere and work with the assumption that a
%unique solution exists for the concrete tensors at hand. 

We now proceed with a decomposition of the classical Yang-Mills energy
momentum tensor obtained from the McLerran-Venugopalan model. We keep the
matching time $\tau_0$ as a parameter, but it is clear that for any practical
applications $\tau_0$ needs to be chosen before the growth of fluctuations could
significantly alter the classical solutions. In addition, for the analytic solutions
discussed here, $\tau_0$ should be within the convergence radius of the
power series in time.

The determination of the viscous fluid fields from Eq.\ (\ref{eq:tmn}) is preferably done 
numerically on the fluid dynamics grid. However it might be instructive
to briefly study one particular case analytically, namely the center of a smooth, event-averaged 
collision at midrapidity. In that case the symmetry arguments collected in Tab.\ 
\ref{tab:parity} apply and the energy momentum tensor in its Cartesian components reads
\begin{widetext}
\begin{equation}
  T^{\mu\nu}_{\mathrm{YM},0} =
  \begin{pmatrix}
    \varepsilon_0-\frac{\tau_0^2}{8}(-2\triangle \epsilon_0 + \delta)  &
    0 & 0 & 0 \\
    0 & \varepsilon_0-\frac{\tau_0^2}{4}(-\triangle \varepsilon_0 +
    \delta-\omega) & 0 & \frac{1}{2} \tau\beta^1  \\
    0  & 0 & \varepsilon_0-\frac{\tau_0^2}{4}(-\triangle \varepsilon_0 +
    \delta+\omega) & 0  \\
    0  & \frac{1}{2} \tau \beta^1 & 0 &
- \varepsilon_0 + \frac{\tau_0^2}{8}(-2\triangle \varepsilon_0 + 3 \delta)
      \end{pmatrix} + \mathcal{O}(\tau^4)  \, .
      \label{eq:tcenter}
\end{equation}
\end{widetext}
As expected from symmetry arguments the velocity $u^\mu=(1,0,0,0)$
corresponds to an eigenvector with rest frame energy $e$ which precisely
resembles the lab frame energy $T^{00}_{\mathrm{YM}}$:
\begin{equation}
  e=\varepsilon_0-\frac{\tau_0^2}{8} \left(-2\triangle \varepsilon_0 +\delta\right) \, .
\end{equation}
One can check that this is the only physically acceptable solution to 
the pertinent eigenvalue problem. From Eq.\ (\ref{eq:ppi}) we know $p+\Pi=e/3$
and the traceless part of Eq.~(\ref{eq:pressure}) yields the shear stress tensor, 
\begin{equation}
\small{  \pi^{\mu\nu} = 
  \begin{pmatrix}
     0 & 0 & 0 & 0 \\ 
     0 & \frac{2}{3}e - \frac{\tau_0^2}{8}\delta+ \frac{\tau_0^2}{4}\omega & 0 & \frac{1}{2} \tau_0\beta^1 \\
     0 & 0 & \frac{2}{3} e -\frac{\tau_0^2}{8} \delta- \frac{\tau_0^2}{4} \omega & 0  \\
     0 & \frac{1}{2} \tau_0\beta^1 & 0 & -\frac{4}{3} e + \frac{\tau_0^2}{4} \delta
  \end{pmatrix}  }\, .
\end{equation}
We can make a few basic observations that turn out to hold numerically for other points
besides the center.
%First, since the classical Yang-Mills theory is conformal,
%matching to a conformal equation of state $p=e/3$ would give vanishing bulk 
%stress. This could be a good approximation for asymptotically large energy 
%densities in QCD. For useful fluid simulations of nuclear collisions one would 
%prefer a more realistic equation of state which breaks 
%conformal symmetry at the applicable energy densities and will lead to finite
%bulk stress. 
We observe that the rapidity-odd flow $\beta^i$ translates directly into
transverse flow of viscous stress $\pi^{0i}$, for $i=1,2$. It does not appear 
in the transverse flow velocity $u^i$. This is an 
immediate consequence of boost-invariance. Boost-symmetry does not permit
rapidity-odd transverse 4-vectors, but rapidity-odd transverse flow components 
are allowed in rank-2 tensors. However, 
%as we will see in the full numerical solutions below, the effects of 
the longitudinal gluon energy flow term $\nabla^i\beta^i$ directly translates into
longitudinal fluid flow $v_3$, although it just happens to vanish at the center point. 
A system without boost-invariance would not have the restriction on the transverse 
flow field. However, for very large collision energies we would expect our observations 
in the boost-invariant case to be a valid approximation around midrapidity.

\section{\label{sec:fluid} Evolution of Initial Angular Momentum in Viscous Fluid Dynamics}

\subsection{Initial Conditions}

Now let us discuss some numerical results for the matching of the classical Yang-Mills
energy tensor to viscous fluid dynamics. As a representative example we continue to study Pb+Pb collisions 
with an impact parameter of $b=6$ fm as before. We set the matching time to $\tau_0 = 0.1$ fm/$c$ and 
choose $Q=1$ GeV in the analytic expressions up to second order in $\tau$.
We set the second order terms $\omega=0=\gamma$ here for simplicity. These terms 
would lead to a pressure asymmetry in the transverse plane which shall be studied
elsewhere. They do not contribute to $dL_2/d\eta$.

\begin{figure}[tb]
\includegraphics[width=\columnwidth]{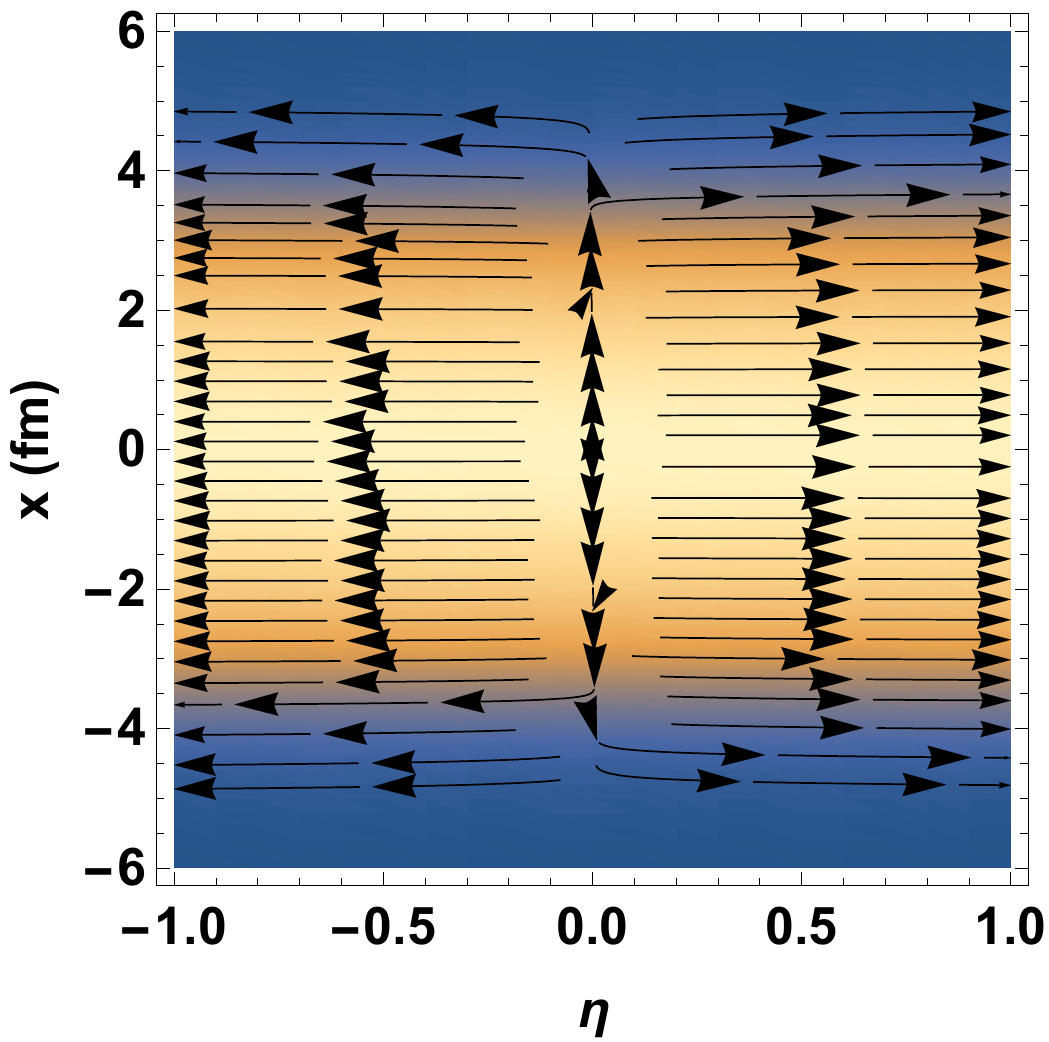}
\caption{\label{fig:fluidflowzx} Fluid flow vector field in the reaction plane,
 $(v_1,v_3)$, together with the local fluid energy density $e$ 
 obtained from the matching at $\tau=0.1$ fm/$c$  as described in the text
 for Pb+Pb collisions at impact parameter $b=6$ fm. The flow is dominated by the Bjorken expansion.
}
\end{figure}

\begin{figure}[tb]
\includegraphics[width=\columnwidth]{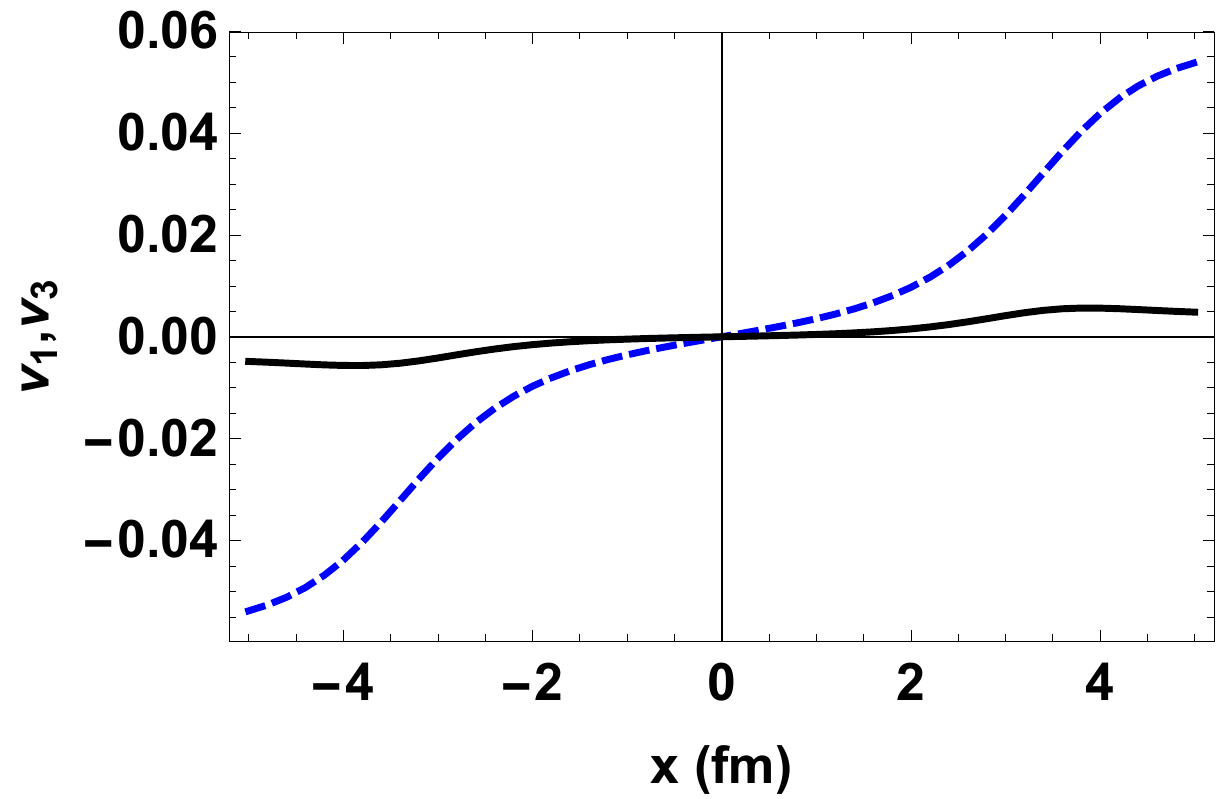}
\caption{\label{fig:fluidfloweta0} Fluid flow vector components $v_1$ (dashed blue line) and $v_3$
(solid black line) as functions of $x$ at $\eta=0$, $y=0$ obtained from the matching at $\tau=0.1$ fm/$c$  
as described in the text for Pb+Pb collisions at impact parameter  $b=6$ fm. Because of the
quadratic time-dependence of the longitudinal shear flow the build-up of $v_3$ is lagging behind
$v_1$.
}
\end{figure}

Fig.\ \ref{fig:fluidflowzx} shows the local energy density $e(\eta,x)$ together with the flow velocity vector $\mathbf{v}(\eta,x)$ in the reaction plane. Predictably the flow is dominated by the
longitudinal Bjorken expansion $v_3 \approx \tanh\eta$. Only around midrapidity can the effect of
transverse flow be competitive. Note that ``naive" boost-invariant fluid dynamics, as it has been
practiced in 2+1D fluid dynamics codes for many years, restricts itself to precisely $v_3 = \tanh \eta$, which does 
not cover all allowed boost-invariant solutions. The longitudinal shear flow is an example for a deviation from this naive scenario. It can be visualized by the behavior of the $z$-component of the fluid velocity, $v_3$, in the plane $\eta=0$.
Fig. \ref{fig:fluidfloweta0} shows $v_1$ and $v_3$ plotted along the $x$-axis at $\eta=0$.
The radial flow at this early time takes maximum values around $\sim 0.05 \, c$ on the surface of the system. The size of $v_3$ at this early time is about an order of magnitude suppressed, but it grows quadratically with time while 
the transverse flow velocity $v_1$ grows linearly.
Interestingly, unlike the energy flow $T^{03}$ there are no nodes in $v_3$ away from the center.
Moreover, the direction of $v_3$ goes in the ``right" direction, i.e.\ along the direction of motion of the nuclei. That means that the shear flow opposite to the nuclear motion, observed in the previous section, must be carried by the dissipative stress tensor.

%\onecolumngrid
\begin{figure*}[tbh]
\begin{center}
\includegraphics[width=0.95\columnwidth]{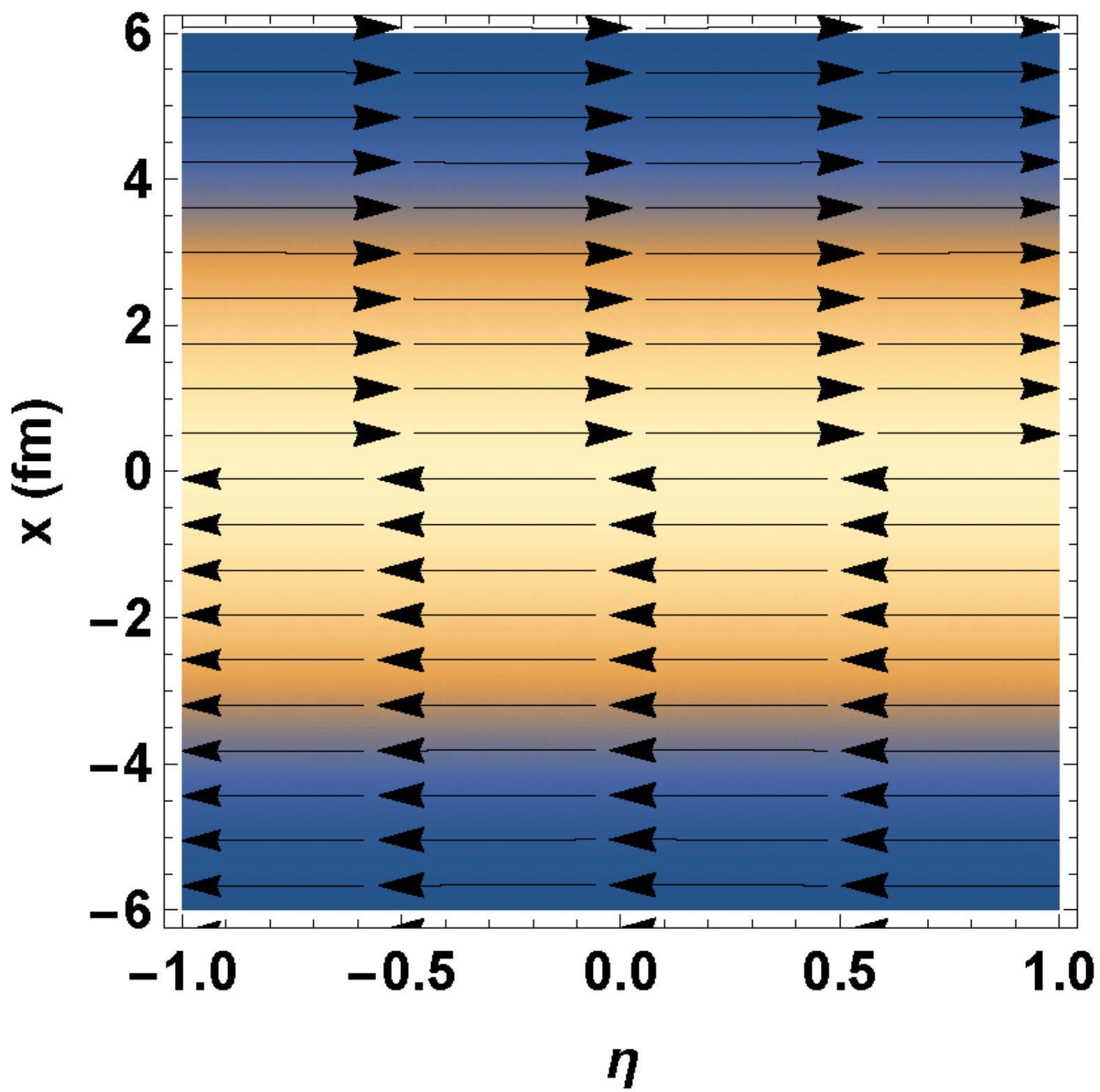}
\includegraphics[width=0.95\columnwidth]{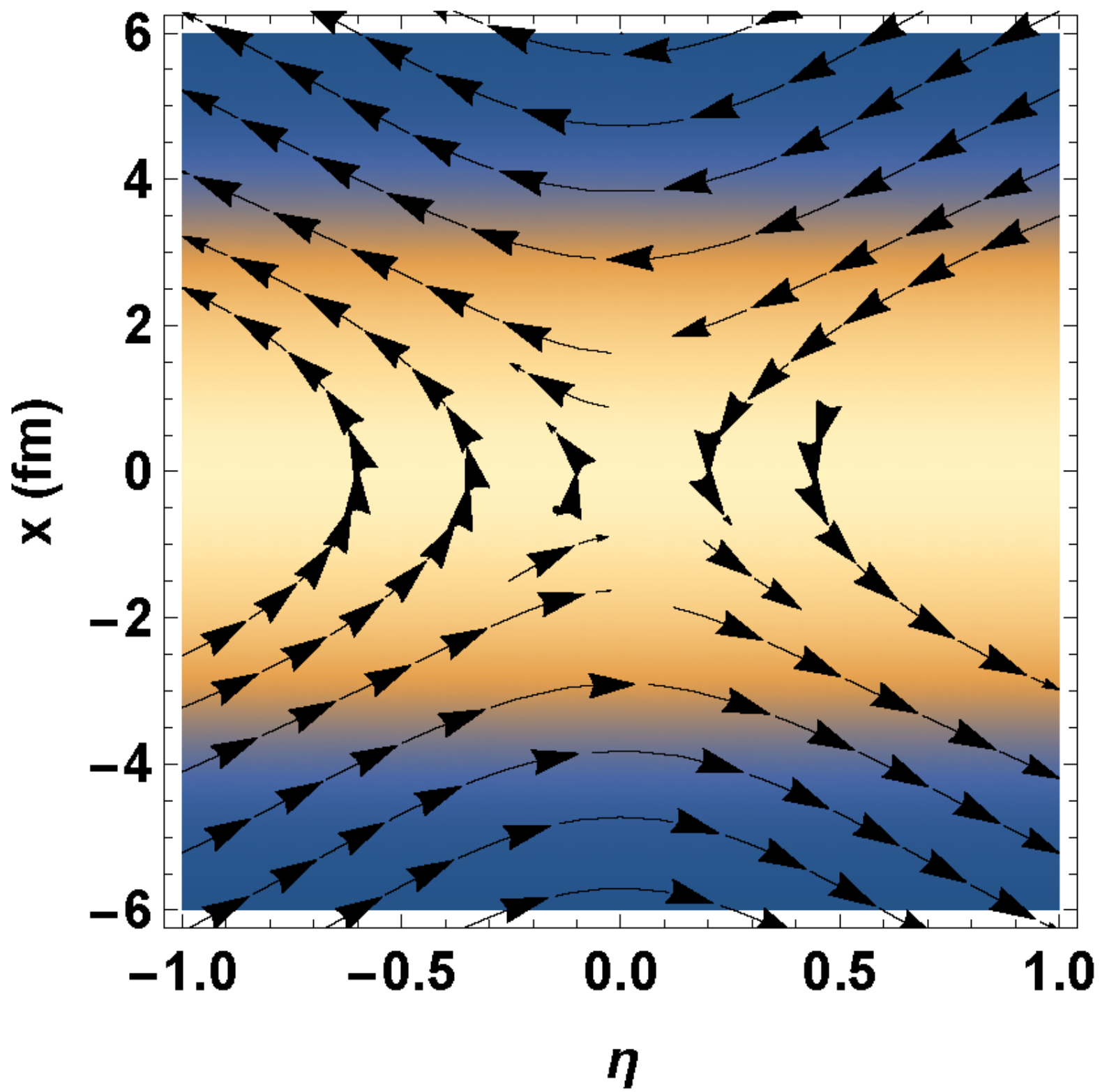}
\end{center}
\caption{\label{fig:decomp} The flow of energy in the reaction plane for Pb+Pb collisions at impact parameter $b=6$ fm, at time $\tau=0.1$ fm/$c$. Left panel: the part of the ideal energy momentum tensor contributing to angular momentum, $(T_{\text{id}}^{01},T_{\text{id}}^{03})_{\text{odd}}$. Right panel: the same for the shear stress tensor, $(\pi^{01},\pi^{03})_{\text{odd}}$. Only the rapidity-even part of longitudinal flow and the rapidity-odd part of transverse flow contribute to angular momentum.
The ideal part exhibits longitudinal shear flow aligned with the motion of the nuclei. The dissipative part carries directed flow aligned with, and longitudinal
shear flow opposite to, the motion of the nuclei.
}
\end{figure*}
%\twocolumngrid

\begin{figure}[tb]
\includegraphics[width=\columnwidth]{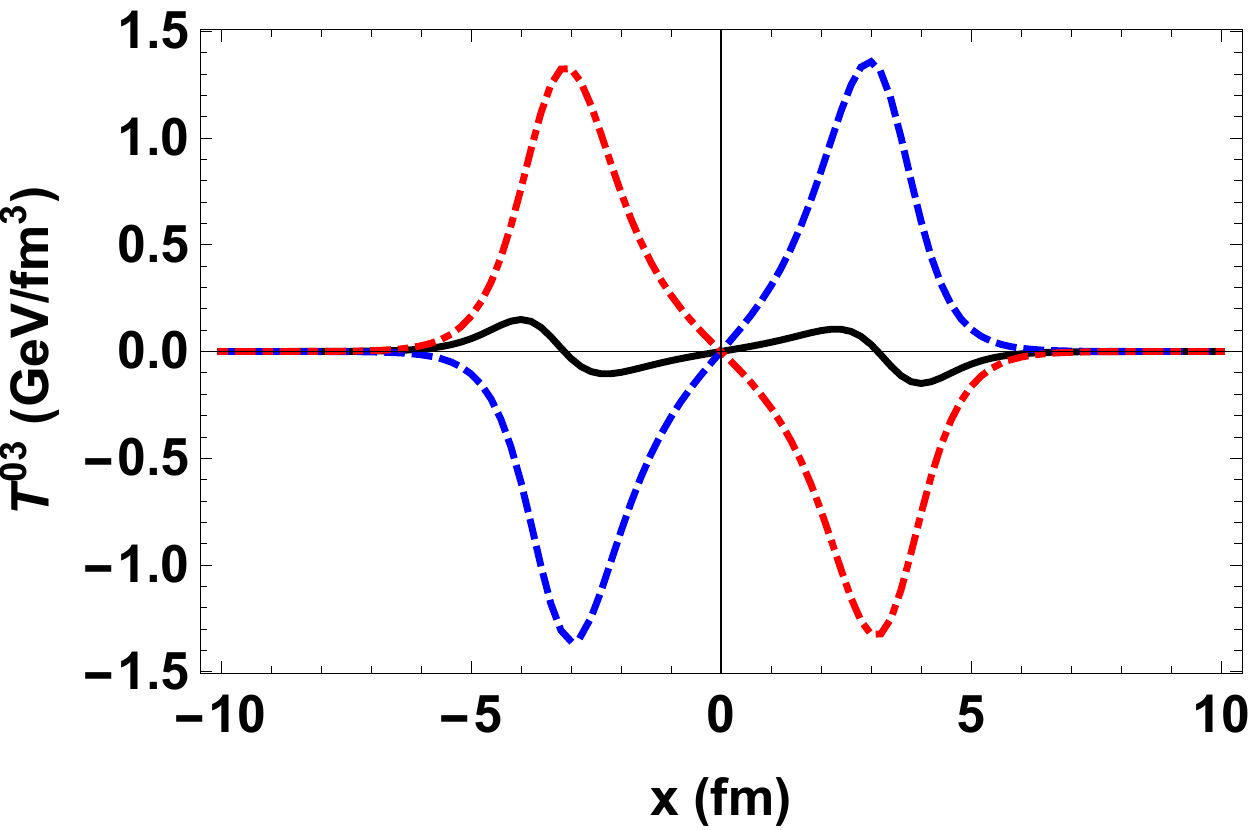}
\caption{\label{fig:t03comp} The longitudinal energy flow $T^{03}$ as a function of $x$ at $\eta=0$ and $y=0$. As in Fig.\ \ref{fig:decomp}
the ideal contribution (dashed blue line) and the dissipative contribution (dot-dashed red line) are shown separately. The total energy flow
(solid black line), with the two additional nodes as discussed in the text, emerges from a non-trivial cancellation of the ideal and dissipative contributions.}
\end{figure}

To separate the two contributions clearly we show in Fig.\ \ref{fig:decomp} separately the ideal part and the dissipative part
of the energy flow in the reaction plane,
\begin{align}
  &\left( T_\mathrm{id}^{01}, T_\mathrm{id}^{03} \right) = \left(e+p\right)  \gamma^2 ( v_1,v_3)  \, &\text{(left panel)}\, ,  \\
   &\left( \pi^{01}, \pi^{03} \right)   \, & \text{(right panel)} \, ,
\end{align}
respectively.
In both cases, for clarity, we again only plot the part of the flow field that carries angular momentum, i.e.\ the rapidity-even part of the 03-component
and the rapidity-odd part of the 01-component, formally defined in the previous section. 
As we expected from Fig.\ \ref{fig:fluidfloweta0} the ideal part of the fluid only exhibits shear flow, whose
direction is consistent with the total system angular momentum. The dissipative stress tensor carries the directed flow in transverse direction,
and a longitudinal shear flow opposing the velocity field. Note that both contributions have to add up to the total energy flow field
and angular momentum discussed in Sec.\ \ref{sec:ymang}. We thus expect the dissipative longitudinal shear flow to be dominant. Indeed,
in Fig.\ \ref{fig:t03comp}, $T^{03}=T^{03}_\mathrm{id} + \pi^{03}$ is plotted along the $x$-axis in the $\eta=0$ plane. We show the ideal part, which basically traces the line of $v_3$ in Fig.\ \ref{fig:fluidfloweta0}, and the dissipative part separately, together with the total.
The ideal and dissipative part have opposite sign and partly cancel each other to give the total energy flow with the two additional nodes
as discussed before.

\begin{figure}[tbh]
\begin{center}
\includegraphics[width=0.95\columnwidth]{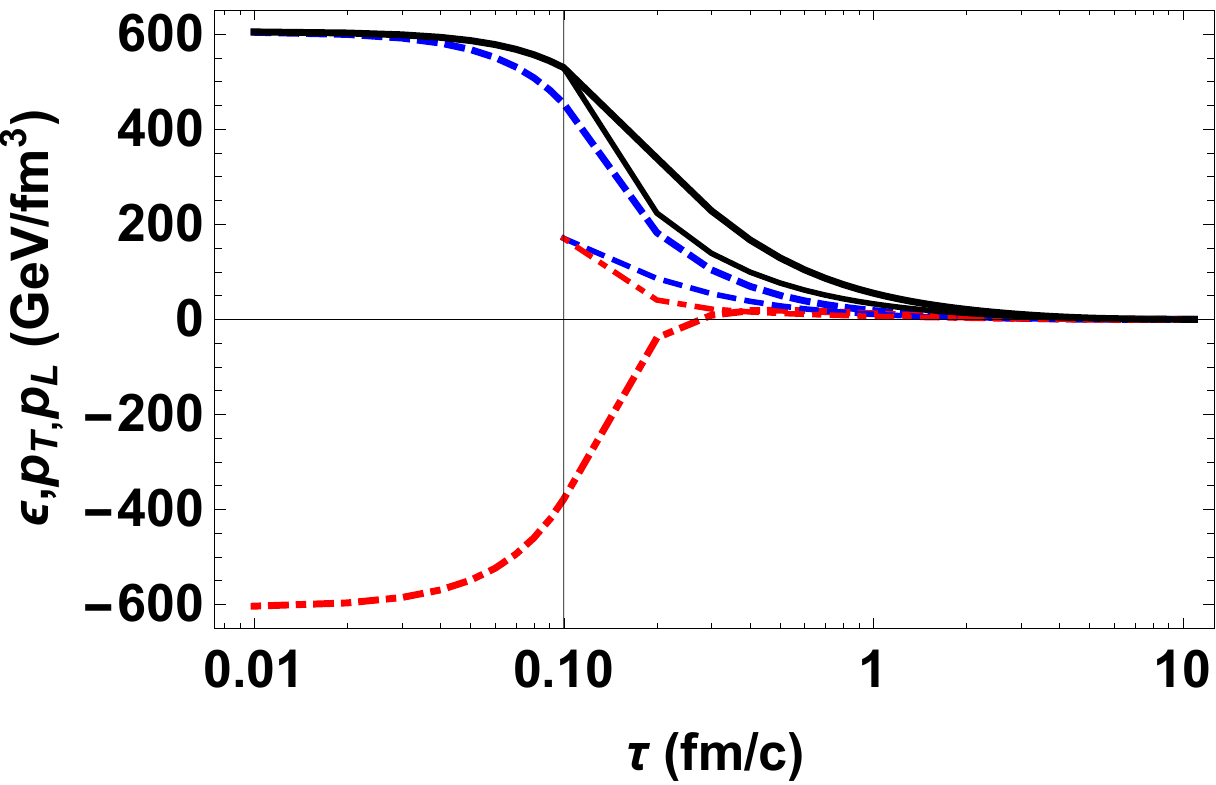}
\end{center}
\caption{\label{fig:eptpl} The time evolution of energy density (solid), transverse pressure (dashed) and longitudinal pressure (dash-dotted line) as a function of time $\tau$ at the center $(x,y,\eta)=(0,0,0)$ of a Pb+Pb collision with $b=6$ fm. For the evolution before $\tau_{0}=0.1$ fm/$c$ we use our classical Yang-Mills formalism. After that time 
the system is evolved with the relativistic viscous hydro code VIRAL in two configurations: (a) dissipative stress is carried over from the classical Yang-Mills simulation to
viscous fluid dynamics (thick lines), (b) dissipative stress is discarded at the beginning of the fluid dynamic simulation. In the latter case longitudinal and transverse pressure are discontinuous at the matching time and the system cools much faster at the center of the collision.
}
\end{figure}

\begin{figure}[tbh]
\begin{center}
\includegraphics[width=0.95\columnwidth]{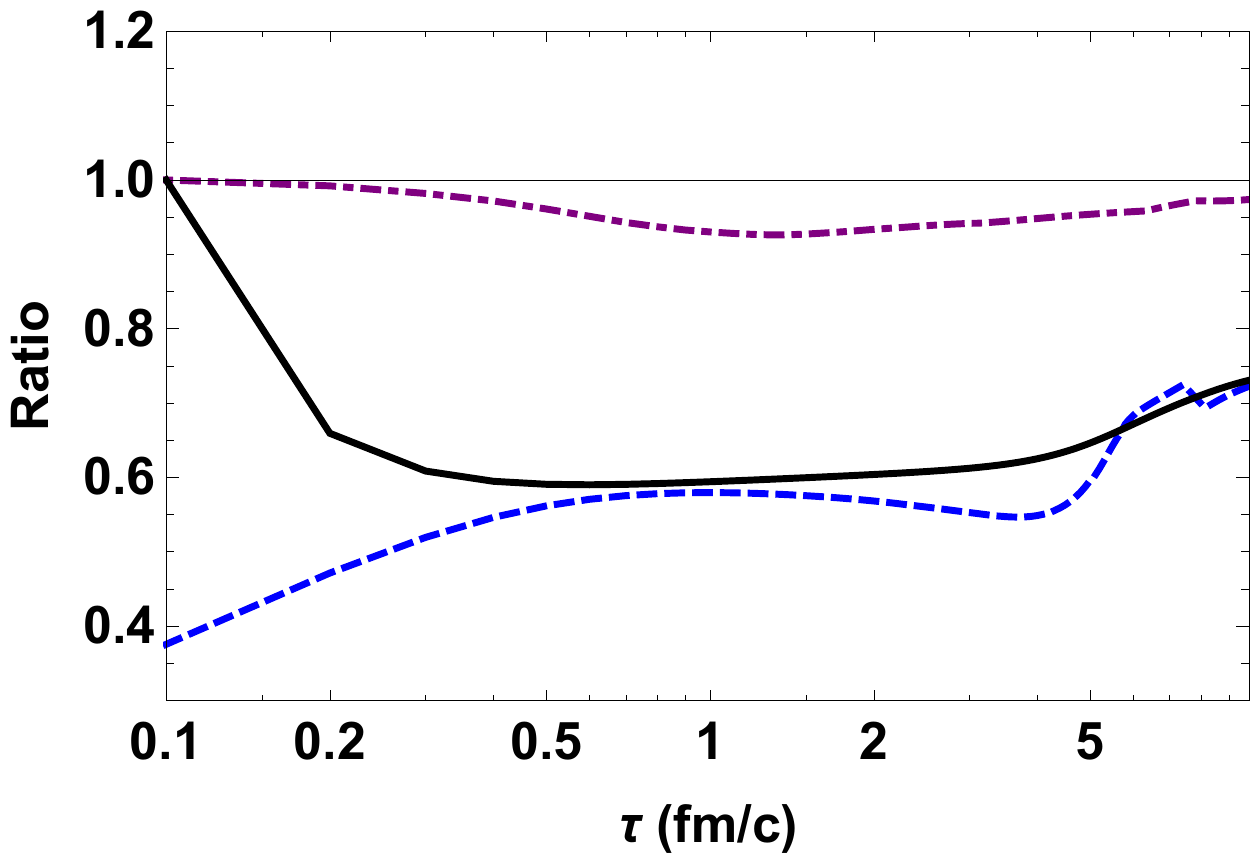}
\end{center}
\caption{\label{fig:dissratios} For the same system as shown in Fig.\ \ref{fig:eptpl} we show the ratio of energy densities in cases (b) and (a) discussed in the text (without and with initial dissipative stress; black solid line) and the ratio of transverse pressures in case (b) and (a) (blue dash-dotted line) as a function of time $\tau$. The energy
density and pressure are calculated in VIRAL fluid dynamics and taken at the center of the collision $(x,y,\eta) = (0,0,0)$. Recall that the absence of dissipative stress here refers to the procedure of discarding the shear and bulk stress found from classical Yang-Mills fields at $\tau=0.1$ fm/$c$. We also show the same ratio for the transverse velocity component $v_1$ measured at a point $(x,y,\eta)=(4 \, \text{fm},0,0)$ away from the center.
Discarding the initial dissipative stress leads to a 30-40\% decrease in energy density and transverse pressure at a given time over the entire time evolution. At the same time
the transverse velocity is reduced by 5-10\%.
}
\end{figure}

We conclude that the decomposition of the angular momentum in terms of fluid fields yields a non-trivial separation of flow and angular 
momentum between the ideal and the dissipative part of the energy momentum tensor. Neglecting the shear stress tensor at the point
of matching could alter the angular momentum and other macroscopic quantities significantly. This is plainly visible in Figs.\ \ref{fig:eptpl} and \ref{fig:dissratios} which
are discussed in detail in the next subsection.

\begin{figure}[tbh]
\begin{center}
\includegraphics[width=0.95\columnwidth]{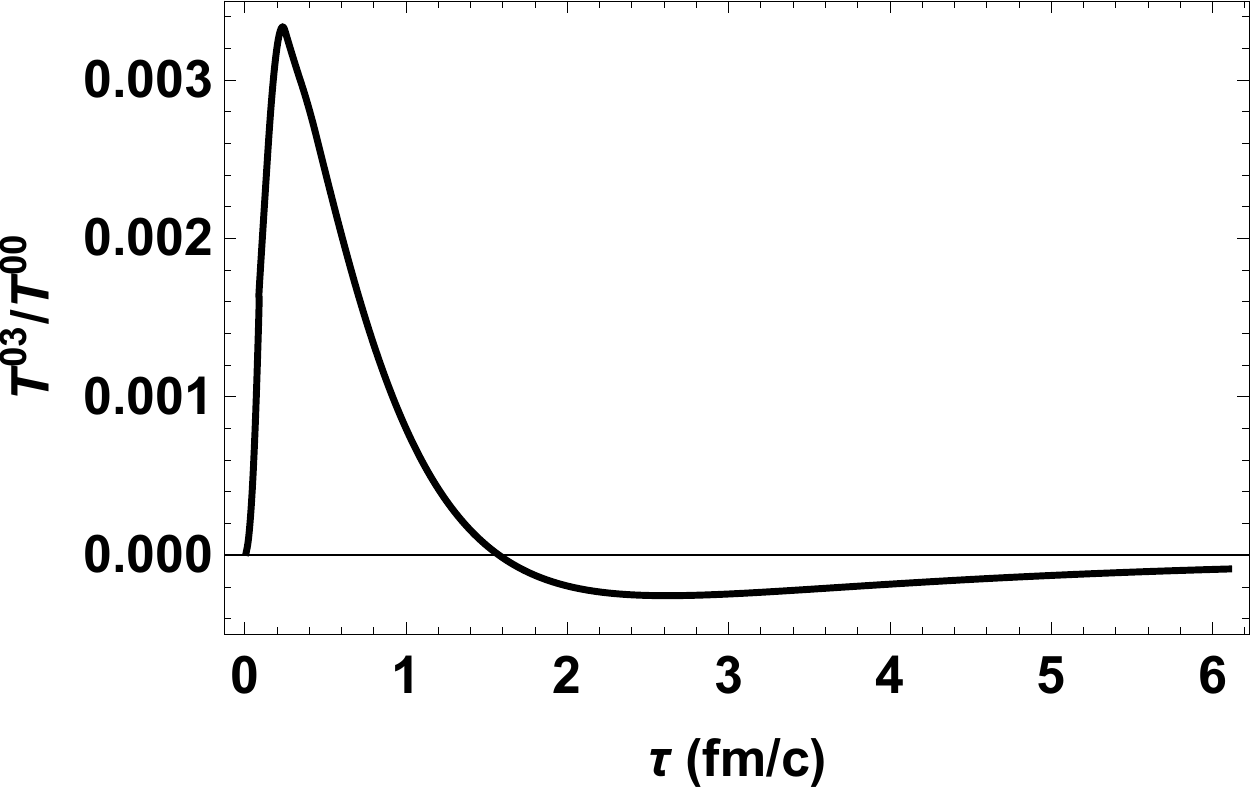}
\end{center}
\caption{\label{fig:flowhydro} The time evolution of the ratio $T^{03}/T^{00}$ as a function of time $\tau$ at the center point $(x,y,\eta)=(-4 \, \text{fm},0,0)$ in a Pb+Pb collision with $b=6$ fm. Again we have used our classical Yang-Mills formalism before time $\tau_{0}=0.1$ fm/$c$ 
and after that time we use the relativistic viscous hydro code VIRAL with initial dissipative stress.
}
\end{figure}

\subsection{Evolution In The Fluid Phase}

One can give simple qualitative arguments about the subsequent evolution of angular momentum in viscous fluid dynamic given initial conditions as discussed in the previous subsection. Here we briefly discuss those arguments and follow up with a numerical simulation using the Texas VIscous RelAtivistic fLuid (VIRAL) code, see Appendix A. We
evolve the system in fluid dynamics with two different initializations: (a) with shear and
bulk stress at the matching time carried over from the classical Yang-Mills phase, and (b) with shear and bulk stress discarded at the matching time. While case (a) seems to be the more physical one, an argument could be made that case (b) is an acceptable approximation. 
Even in case (b) bulk and shear stress build up with time, driven by the gradients in the system, as long as bulk and shear viscosities are finite. In fact, interestingly we find that
the Navier-Stokes values $\pi^{\mu\nu}_{\mathrm{NS}}= 2\eta d^{\langle \mu} u^{\nu\rangle}$ for the shear stress, 
calculated from the initial flow field at the matching time, are remarkably similar to the true values $\pi^{\mu\nu}$ extracted from the classical Yang-Mills system. To be more precise, components of the two tensors \emph{qualitatively} have the same functional dependence on position, and the same sign. One could roughly write $\pi^{\mu\nu} \approx C \pi^{\mu\nu}_{\mathrm{NS}}$ with $C>0$. $C$ varies from component to component but is generally of order one if minimal viscosities are used in the Navier-Stokes approximation.
Here we need to define
\begin{equation}
  d^{\langle \mu} u^{\nu\rangle} = \frac{1}{2} \left( d^\mu u^\nu + d^\nu u^\mu   
  \right) -\frac{1}{3}\Delta^{\mu\nu} d_\mu u^\mu   \, ,
\end{equation}
where $\Delta^{\mu\nu} = g^{\mu\nu}-u^\mu u^\nu$ is the usual projection operator orthogonal to the flow
velocity, and $d^\mu = \Delta^\mu_\nu \partial^\nu$ \cite{Muronga:2004sf}.
The argument in favor of (b) would be that the second order corrections in viscous fluid dynamics, 
$\pi^{\mu\nu} - \pi^{\mu\nu}_{\mathrm{NS}}$, should be small. In practice, this difference matters greatly at early matching times.

Quantitatively there are sizable differences between running VIRAL for scenarios
(a) and (b). Fig.\ \ref{fig:eptpl} shows the evolution of energy density, transverse pressure and longitudinal pressure at the center of the collision $(x,y,\eta)=(0,0,0)$ as a function of time. Note that the lab frame and local fluid rest frame coincide at the point shown and we plot both the evolution in the classical Yang-Mills phase ($\tau<0.1$ fm) and the viscous fluid phase ($\tau>0.1$ fm) in case (a). Per our matching procedure energy densities and pressures evolve smoothly across the matching time
$\tau_0$. The VIRAL code is stable although the longitudinal pressure is still negative at initialization. Of course, higher order gradient corrections to second order viscous fluid dynamics, not included here, could potentially be large. For $\tau>\tau_0$
we also show the results of initializing VIRAL in case (b), when the initial dissipative stress is discarded. The
energy density and pressure are now permitted to be discontinuous and indeed exhibit large jumps at $\tau_0$. The energy density is
continuous here only due to the fact that at the chosen point at the center the local rest frame is also the lab frame which 
enforces $\pi^{00} = 0$ by construction. We notice that the system cools much faster and at the same time has a lower transverse pressure and larger longitudinal pressure compared to case (a).

The comparison is continued in Fig.\ \ref{fig:dissratios} where the ratio of cases (b) and (a)
is plotted for the energy density and transverse pressure at the center, and for the transverse velocity at a point 5 fm away from the
center. The faster cooling and reduced transverse pressure in case (b) are clearly visible and persist to very late times. The effect on the transverse velocity is not as pronounced, case (b) lags behind case (a) by about 5-10\%. The discrepancy might decrease if later matching times
are chosen, although the validity of the classical Yang-Mills approach becomes doubtful. 
%Here, we are limited in our reach in times by only studying the Yang-Mills solution up to second order in 
%time. 
In any case, keeping the dissipative stress when initializing the fluid dynamics is the preferred method, despite
uncertainties coming from large gradients. 
 %our arguments in the next section are mainly for that case, although they qualitatively also hold for case (b).

Let us discuss the fate of angular momentum. The first observation is that shear viscosity will dampen the 
longitudinal shear flow shown in the left panel of Fig.\ \ref{fig:decomp}. At the same time the shear stress tensor 
will relax to its Navier-Stokes value $\pi_{\mathrm{NS}}^{\mu\nu}$. 
At the beginning of the fluid dynamic evolution the angular momentum in the shear stress 
tensor is carried by the rapidity-odd part of $\pi^{01}$ and the rapidity-even part of $\pi^{03}$. It is easy to 
check that boost-symmetry ensures that the corresponding Navier-Stokes quantities, the rapidity-odd part of 
$\pi^{01}_{\mathrm{NS}}$ and the rapidity-even part of $\pi^{03}_{\mathrm{NS}}$ vanish. Thus the flow 
carrying angular momentum on the right hand side of Fig.\ \ref{fig:decomp} will be damped.
%on a time scale given by the shear relaxation time.

In summary, from simple arguments we expect that longitudinal shear flow and the angular momentum density at midrapidity tend to die out in the viscous fluid phase of a system with exact boost-invariance. 
%Of course, the equations of motion respect local angular momentum conservation, but the total angular momentum in the system is not finite. 
The longitudinal flow of angular momentum, $dN_2/d\eta$ in early fluid dynamics is thus reversed compared to the
Yang-Mills phase.
Our expectations are confirmed by running the VIRAL fluid code.
Fig.\ \ref{fig:flowhydro} presents a typical example for the time evolution of the longitudinal shear flow. The ratio $T^{03}/T^{00}$ at a point at midrapidity but away from the center of the transverse plane is depicted as a 
function of time in case (a). The matching from classical Yang-Mills evolution to fluid dynamics is smooth but leads to a rapid change of the time derivative which dissipates the shear flow. Thus, even though matching procedure (a)
ensures continuity and macroscopic conservation
laws, we have an explicit example of a non-smooth time evolution.
This is not unexpected as important microscopic physics is missing when Gauss' Law is 
instantaneously replaced by a dissipative law at time $\tau_0$. 
%A more realistic picture would involve a smooth transition of the equations of motion.

Let us for a moment look at the larger picture. Realistic systems obey boost-invariance only around mid-rapidity
and the total angular momentum of the system is finite and constant. The arguments given above will still hold
approximately around midrapidity, but will fail away from midrapidity. Under realistic conditions the flow field can support directed flow which corresponds to a rotational motion of the system. Indeed, experimental measurements see
that quantities directly sensitive to angular momentum, e.g.\ the directed flow of particles \cite{Adamczyk:2014ipa}, 
and the polarization of $\Lambda$ hyperons \cite{STARLambda} notably decrease at midrapidity with increasing 
collision energy. They are suppressed by the increasingly well realized boost-invariance. A quantitative study of these effects will require a description of the initial state with a realistic rapidity profile, and its dependence on the
collision energy. Lastly, we mention that boost-invariance prevents fluctuations of the fluid dynamic system in longitudinal direction. Such fluctuations can lead to Kelvin-Helmholtz instabilities and the formation of smaller vortices in the fluid phase
\cite{Csernai:2011qq,Okamoto:2017ukz}.

\section{Conclusions}

In summary, we have estimated the event-averaged angular momentum in nuclear collisions as a function of space-time rapidity,
within in the McLerran-Venugopalan model of classical gluon fields, at very early times $\lesssim 1/Q_s$.
The results can serve as estimates around midrapidity for realistic collisions at top RHIC energies and higher, where
boost-invariance holds approximately. We find that $dL_2/d\eta$ peaks in mid-central collisions for $b\approx 5$ fm
for collisions of Pb nuclei. From Eq.\ (\ref{eq:l2est}) we read off that $dL_2/d\eta \approx \frac{1}{2} 
R_A Q_s^{-3} \bar\varepsilon_0 $ at time $\tau \approx 1/Q_s$ at midrapidity. 
The build-up of angular momentum at midrapidity is driven by the QCD version of Gauss' Law. The angular momentum can be visualized as a vortex 
in the gluon energy flow in the reaction plane, with an opposing longitudinal shear flow of energy. The net effect is angular 
momentum $dL_2/d\eta$ opposite to the total primordial angular momentum. 

We have discussed the direct matching of classical gluon fields at early time to relativistic fluid dynamics, following 
established precedence. We have derived some fundamental statements both regarding matching to ideal and
second order viscous fluid dynamics. If conservation laws for energy, momentum and angular momentum are used
as the guiding principle the projection of the energy momentum tensor perpendicular to the matching hypersurface
is always continuous. In ideal fluid dynamics other components might be discontinuous while viscous fluid dynamics allows
for a smooth energy momentum tensor across the matching. However, dissipative stress needs to be kept in the fluid
initial conditions to ensure macroscopic conservation laws and continuous energy momentum and angular momentum tensors.
%Failure to do so will violate conservation laws and lead to quantitative differences in the fireball evolution. We did not 
%explore this fact further as the correct choice seems clear. 
As a caveat we have pointed out that even if dissipative stress is kept some equations of motion are discontinuous 
across the matching surface. Shear in the longitudinal energy flow is a relevant example, with the time derivative changing 
sign abruptly around the matching time. Thus the total amount of angular momentum built up at midrapidity depends critically on 
the matching time chosen. A more reliable estimate would have to take into account how the macroscopic mechanism of 
Gauss' Law is broken by the onset of decoherence of the classical fields. We will also mention once more that
second order viscous fluid dynamics will evolve systems with large initial shear stress with significant uncertainties which 
can be checked by comparing to anisotropic fluid codes.

In the subsequent fluid dynamic evolution angular momentum is initially carried by the longitudinal shear flow and by 
directed flow in the shear stress tensor. This is dictated by boost-invariance. Both modes are damped in viscous 
fluid dynamics. As a result the angular momentum around midrapidity decays quickly with time. This result is consistent 
with small or vanishing directed flow and particle polarization seen at top RHIC energies. 

Although the final angular momentum is small, there are several conclusions we can draw.
First, we have clarified the mechanism through which angular momentum can be transported to midrapitity in the initial
CGC phase. Secondly, we have described the very simple mechanisms through which the angular momentum dissipates
in a boost-invariant fluid system. We emphasize once more that angular momentum is conserved throughout the entire calculation, however the total amount of angular momentum in the boost-invariant system is not well defined.
Our study can serve as a starting point for calculations at lower collision energies or at larger rapidities, e.g.\ if boost-invariance is explicitly broken with additional assumptions in the fluid dynamic phase. This would allow for 
directed flow at larger rapidities. In the future, event-by-event calculations should also study the effect of fluctuations.

\begin{figure}[tbh]
\begin{center}
\includegraphics[width=0.95\columnwidth]{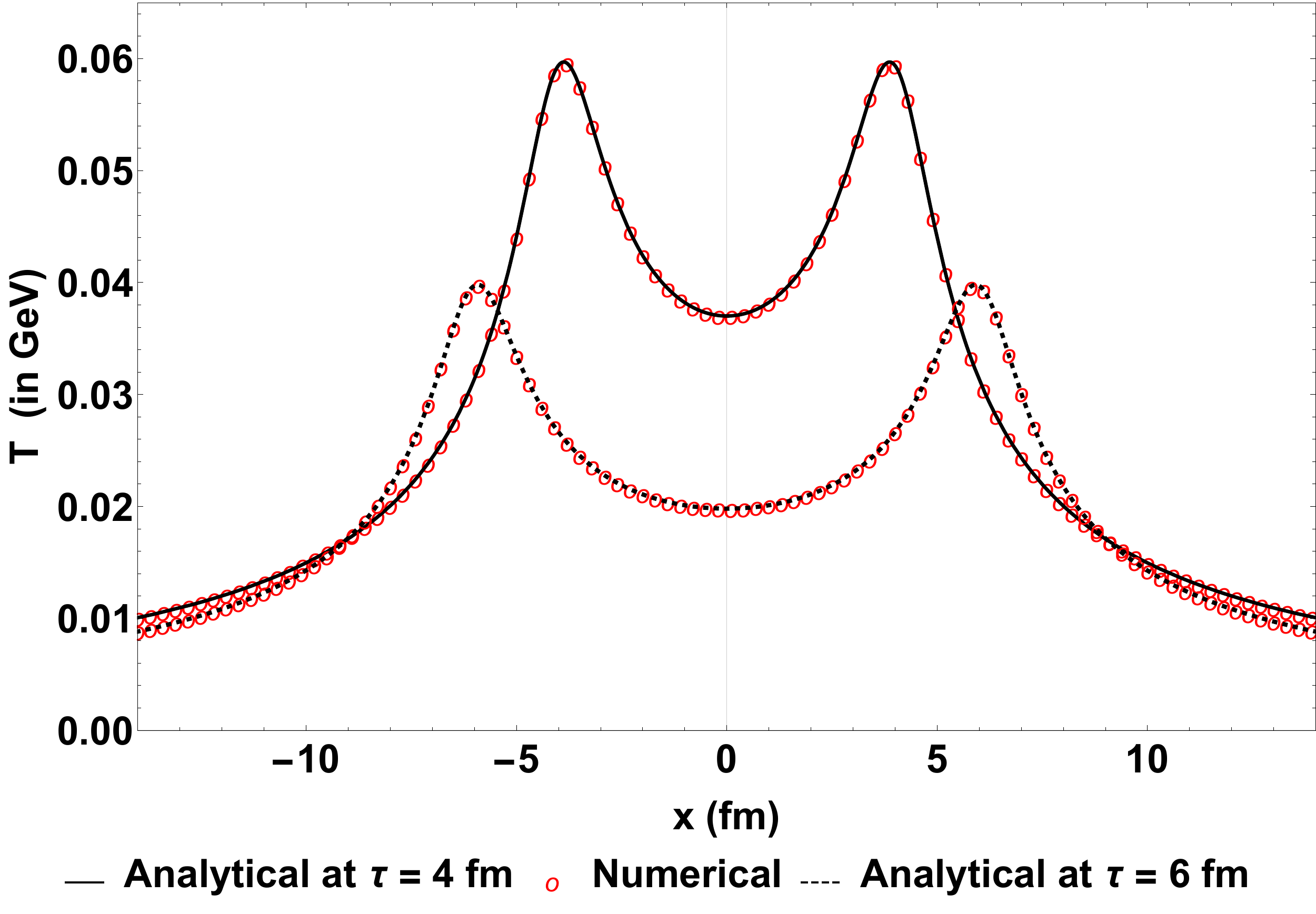}
\includegraphics[width=0.95\columnwidth]{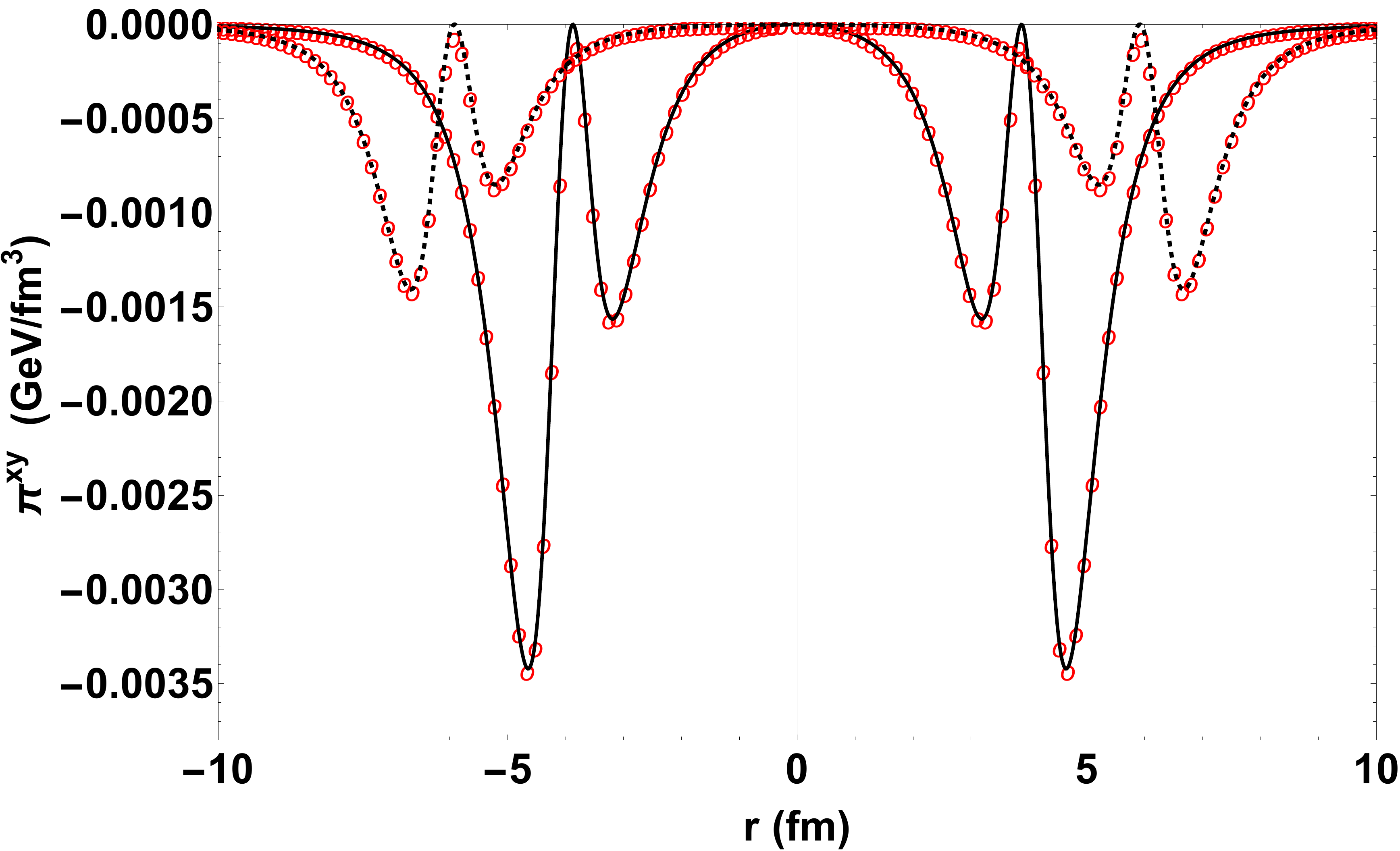}
\end{center}
\caption{\label{fig:gubser} Analytic results (black solid lines) and results obtained with the VIRAL code (red circles) for the cold plasma limit problem  described in \cite{Marrochio:2013wla}. We show the temperature profile along the 
  $x$-axis (upper panel) and the shear stress component $\pi^{xy}$ along a line $x=y$ (lower panel) at midrapidity for times 4 fm/$c$     
  and 6 fm/$c$. The system is initialized at time 1 fm/$c$.
}
\end{figure}

%\begin{appendix}

\appendix*

\section{The VIRAL Fluid Code}

For this work we have utilized the VIscous RelAtivistic fLuid (VIRAL) code developed locally. We will introduce details of this
code in more detail elsewhere, but we want to briefly summarize its technical specifications and abilities. 
VIRAL solves the equations of motion of second order viscous relativistic fluid dynamics in 3+1 dimensions \cite{Israel:1979wp,Muronga:2004sf,Schenke:2010nt,Karpenko:2013wva}, written as conservation laws with source terms. Only five independent components of the shear 
stress tensor are treated as dynamical quantities, the remaining components are reconstructed from constraints. The conservation
laws are solved through the improved fluxes suggested by Kurganov and Tadmor \cite{KurTad:2000}, using 5th 
order WENO (weighted essentially non-oscillatory) spatial derivatives \cite{weno}. The time integration is carried out with a 3rd order TVD 
(total variation diminishing) Runge-Kutta scheme \cite{tvdrk}. The code is written in C++ with built in MPI capabilities. 
For the calculations in Sec.\ \ref{sec:fluid} we have used the equation of state s95p-PCE165-v0 \cite{Huovinen:2009yb,eos:huovinen}, and constant specific shear viscosity $\eta/s=1/4\pi$ and specific bulk viscosity $\zeta/s=0.01/4\pi$.

In order to offer a quick but non-trivial check of the code we present results of a test desribed in \cite{Marrochio:2013wla} based on the work
by Gubser et al. \cite{Gubser:2010ze,Gubser:2010ui}. In Fig.\ \ref{fig:gubser} we show analytic results for the temperature and one 
component of the shear stress tensor in the cold plasma limit discussed in \cite{Marrochio:2013wla}, together with results obtained with 
VIRAL for the same setup. The VIRAL results are virtually indistinguishable from the analytic results despite the steep gradients.
For this test the code was run with a conformal equation of state $e=3p$ and $\eta/s=0.2$. For details of the problem solved here 
we refer the reader to \cite{Marrochio:2013wla}.

%\end{appendix}

%\subsection{}
%\subsubsection{}

% If in two-column mode, this environment will change to single-column
% format so that long equations can be displayed. Use
%\begin{widetext}
%\end{widetext}

% entire page. There is no need to do explicit centering.

% \begin{figure}
% \end{figure}

% Surround figure environment with turnpage environment for landscape
% figure
% \begin{figure}
% \end{figure}

%\appendix
%\section{}

\begin{acknowledgments}
RJF would like to thank Charles Gale and Sangyong Jeon for their hospitality at McGill University where part of this work was carried out. We
are grateful to Joseph Kapusta for useful discussions. Important resources were provided by Texas A\&M High Performance Research Computing.
This work was supported by the US National Science Foundation under award no.\ 1516590 and award no.\ 1550221. 
GC acknowledges support by the Department of Energy under grant no.\ DE-FG02-87ER40371.
\end{acknowledgments}


\end{document}